\documentclass[11pt,a4paper]{article}

\usepackage{amsmath,amsfonts,amssymb,mathrsfs,mathabx}
\usepackage{dsfont}
\usepackage{cite}
\usepackage[dvips]{graphicx}
\usepackage{tikz}
\usetikzlibrary{positioning,shapes,arrows}

\usepackage{txfonts}

\usepackage{hyperref}

\hypersetup{
    unicode=true,          
    pdftoolbar=false,        
    pdfmenubar=false,        
    pdffitwindow=false,     
    pdfstartview={FitH},    
    pdftitle={Generalized Mathieu equation},    
    pdfauthor={M. Piatek, A. R. Pietrykowski},     
    pdfsubject={Integrable systems},   
    pdfcreator={M. Piatek, A. R. Pietrykowski},   
    pdfproducer={M. Piatek, A. R. Pietrykowski}, 
    pdfkeywords={Integrable systems} {Supersymetry}
                {Seiberg-Witten theory} {AGT correspondence} {Mathieu equation}, 
    pdfnewwindow=true,      
    colorlinks=true,       
    linkcolor=black,          
    citecolor=blue,        
    filecolor=blue,      
    urlcolor=blue           
}

\def \a {\alpha}

\def \d {\delta}
\def \e {\epsilon}

\def \vf{\varphi}

\def \ka {\kappa}
\def \la {\lambda}

\def \r {\rho}

\def \m{\mu}
\def \n{\nu}

\def \La{\Lambda}
\def \dl{\Delta}
\def \pt{\partial}
\def \ex{\mbox{e}}

.tfm scaled 1000 
.tfm scaled 1000 
.tfm scaled 1000 

\newcommand \ord[1]{\mathcal{O}({#1})}
\newcommand \cbr[1]{\left({#1}\right)}
\newcommand \sbr[1]{\left[{#1}\right]}

\newcommand \bra[1]{\langle {#1} |}
\newcommand \ket[1]{|{#1} \rangle}
\newcommand \bracket[2]{\langle {#1} | {#2} \rangle}

\numberwithin{equation}{section}
\def\baselinestretch{1.2}

\newcommand{\preprintsize}{
     \topmargin= 0 cm \headsep=0.1cm
     \oddsidemargin= 0cm
      \evensidemargin= 0cm  
      \textheight = 23truecm \textwidth=16truecm      %
}

\preprintsize

\begin{document}

\begin{center}
{\LARGE Classical irregular blocks, Hill's equation 
\\[8pt]
and 
PT--symmetric periodic complex potentials}
\end{center}
\bigskip
\begin{center}
{\large\textsc{Marcin Piatek}$^{\,a,\,c,\;}$}
\footnote{\href{mailto:piatek@fermi.fiz.univ.szczecin.pl}{e-mail: piatek@fermi.fiz.univ.szczecin.pl}}
\hskip 1.0cm
{\large\textsc{Artur R. Pietrykowski}$^{\,b,\,c,\;}$}
\footnote{\href{mailto:pietrie@theor.jinr.ru}{e-mail: pietrie@theor.jinr.ru}}

\vskip 4mm
${}^{a}$
Institute of Physics, University of Szczecin\\
ul. Wielkopolska 15, 70-451 Szczecin, Poland

\vskip 4mm
${}^{b}$
Institute of Theoretical Physics\\
University of Wroc{\l}aw\\
pl. M. Borna, 950-204 Wroc{\l}aw, Poland

\vskip 4mm
${}^{c}$
Bogoliubov Laboratory of Theoretical Physics,\\
Joint Institute for Nuclear Research, 141980 Dubna, Russia
\end{center}

\vskip .5cm

\begin{abstract}
\noindent
The Schr\"{o}dinger eigenvalue problems for 
the Whittaker--Hill potential $Q_{2}(x)=\tfrac{1}{2} h^2\cos4x +$\\ $4h\mu\cos2x$ and
the periodic complex potential $Q_{1}(x)=\tfrac{1}{4}h^2{\rm e}^{-4ix}+2h^2\cos2x$
are studied using their realizations in two-dimensional conformal field theory (2dCFT). 
It is shown that for the weak coupling (small) $h\in\mathbb{R}$
and non-integer Floquet parameter $\nu\notin\mathbb{Z}$ 
spectra of hamiltonians ${\cal H}_{i}\!=\!-{\rm d}^2/{\rm d}x^2 + Q_{i}(x)$, $i=1,2$ 
and corresponding two linearly independent eigenfunctions
are given by the classical limit of the ``single flavor'' and ``two flavors'' 
($N_f=1,2$) irregular conformal blocks. It is known that complex non-hermitian
hamiltonians which are {\sf PT}-symmetric ( = invariant under simultaneous  parity {\sf P} and time 
reversal {\sf T} transformations) can have real eigenvalues.
The hamiltonian  ${\cal H}_{1}$ is {\sf PT}-symmetric for $h,x\in\mathbb{R}$. It
is found that ${\cal H}_{1}$ has a real spectrum in the weak coupling region 
for $\nu\in\mathbb{R}\setminus\mathbb{Z}$. This fact in an elementary way
follows from a definition of the $N_f=1$ classical irregular block.
Thus, ${\cal H}_{1}$ can serve as yet another new model 
for testing postulates of {\sf PT}-symmetric quantum mechanics.
\end{abstract}

\newpage
{\small \hrule \tableofcontents \vskip .5cm\hrule}

\section{Introduction}
\label{intro}
The Hill's equation is a second-order linear ordinary differential equation 
of the (Schr\"{o}dinger-like) form:
\begin{equation}
\label{HillEq}
-\psi''(x)+Q(x)\psi(x)\;=\;\lambda\,\psi(x),
\quad
x\in\mathbb{R}/\pi \mathbb{Z},
\end{equation}
where the function $Q(x)$ may generically be complex valued and it is assumed to be of bounded variation and 
periodic, with the base period $\pi$, i.e., $Q(x+\pi) = Q(x)$ \cite{MagnusWinkler}. The two fundamental solutions 
to eq. \eqref{HillEq} can be cast into the form that reveals their periodicity which is
possible due to the Floquet's Theorem (for more details see appendix \ref{sec:App_A}). 
Moreover, the Oscillation Theorem asserts that the (real) spectrum of the Hill's operator 
with \emph{real} $Q(x)$ reveals a band structure \cite{MagnusWinkler}. The bands are
open sets in the positive real line separated by gaps. The solutions that
depend on spectral values that fall in to the bands have bounded variation in opposite to
those whose spectral values fall into gaps.

The simplest, but very important special cases of the Hill equation are
the Mathieu equation \cite{Mathieu:1868} which assumes 
\begin{equation}\label{Mathieu}
Q(x) \;=\; 2h^2\cos2x,
\end{equation}
and the Whittaker--Hill equation \cite{MagnusWinkler} 
for which\footnote{In the present work: ${\mathrm A}=\frac{1}{2}h^2$, ${\mathrm B}=4h\mu$.}
\begin{equation}\label{WH}
Q(x) \;=\; \mathrm{A}\cos4x+\mathrm{B}\cos2x.
\end{equation}

Eqs.~(\ref{HillEq})--(\ref{Mathieu}) and (\ref{HillEq})--(\ref{WH}) 
occur in a broad spectrum of physical problems. 
Their solutions proved useful in many fields of engineering 
\cite{McLachlan:1947,Paul:1990zz}, quantum chemistry \cite{RA}
and pure physics ranging from some topologically non-trivial gauge theories \cite{Cho:2003rp} 
to cosmology \cite{Lachapelle,Lachapelle:2008sy} and D-brane physics 
\cite{Gubser:1998iu,Manvelyan:2000yv,Park:2000iy}.
Obviously, Schr\"{o}dinger operators with periodic potentials are of special importance
in solid state physics.
Potentials (\ref{Mathieu}) and (\ref{WH}) are real, hence corresponding quantum--mechanical (QM)
hamiltonians are hermitian and have real spectra. Recall, that also some complex potentials have
applications in quantum physics, for instance, in nuclear theory \cite{Srivastava}. 
Another especially interesting complex potentials are those which yield ${\sf PT}$--symmetric
QM hamiltonians.\footnote{The hamiltonian ${\cal H}$ is ${\sf PT}$--symmetric if it satisfies
${\cal H}=({\sf PT}){\cal H}({\sf PT})$,
where the symbol ${\sf P}$ represents the space reflection operator (parity operator)
and ${\sf T}$ stands for the time reversal operator, cf.~\cite{Bender:2007nj}. 
The effect of ${\sf P}$ and
${\sf T}$ on the QM coordinate operator $\hat x$ and the momentum operator $\hat p$
is as follows:
\begin{eqnarray*}
{\sf P}\,\hat x\,{\sf P}=-\hat x, && {\sf P}\,\hat p\,{\sf P}=-\hat p,
\\
{\sf T}\,\hat x\,{\sf T}=\hat x, && {\sf T}\,\hat p\,{\sf T}=-\hat p.
\end{eqnarray*}
In the Schr\"{o}dinger 
eigenvalue problem ${\cal H}\psi=\lambda\psi$ we have $\hat x\mapsto x$ and $\hat p\mapsto -i\frac{d}{dx}$.
The parity operator ${\sf P}$ is a linear operator and that it leaves invariant the 
commutation relation $[\hat x, \hat p]=i\hbar$. The same is true for ${\sf T}$ if 
it is assumed that ${\sf T}i{\sf T}=-i$.
Since ${\sf P}$ and ${\sf T}$ are reflection operators, their squares are the unit operator:
${\sf P}^2={\sf T}^2={\bf 1}$. Finally, it follows that $\left[{\sf P}, {\sf T}\right]=0$.} 
Case studies show that ${\sf PT}$--symmetric hamiltonians may have 
real spectra.\footnote{Precisely, the eigenvalues of a particular 
{\sf PT}--symmetric hamiltonian are real if every eigenfunction of a {\sf PT}--symmetric hamiltonian
is also an eigenfunction of the {\sf PT} operator, cf.~\cite{Bender:2007nj}.} 
In the present work we will consider in particular the eigenvalue problem (\ref{HillEq}) with the complex 
potential:
\begin{equation}\label{complexQ}
Q(x)\;=\;\frac{1}{4}h^2 {\rm e}^{-4ix} + 2h^2\cos2x
\end{equation}
which is obviously {\sf PT}--symmetric for $h, x\in\mathbb{R}$.\footnote{{\sf PT}--symmetric
complex potentials satisfy $\overline{Q(x)} = Q(-x)$.}

\begin{figure}[t]
{
\def\baselinestretch{.8}
\tikzstyle{abstract}=[rectangle, draw=black, rounded corners, text width=3.8cm,align=flush center,inner sep=2.5pt]
\tikzstyle{comment}=[rectangle, draw=black, rounded corners, 
        text centered, anchor=north, text=red, text width=3cm]
\tikzstyle{darrow}=[>=stealth,very thick,<->]
\tikzstyle{sarrow}=[>=stealth,very thick,->]
\tikzstyle{desc} = [text width=3cm,align=#1,midway]
\begin{tikzpicture}[node distance=1.2cm]
    \node (2dCFT) [abstract, rectangle split, rectangle split parts=2]
        {\small
            \textbf{2d CFT}
            \nodepart{second}\footnotesize\textcolor{red}{Virasoro conformal\\ blocks}
        };
        \node (4dN2SYM) [abstract, rectangle split, rectangle split parts=2,right=of 2dCFT]
        {\small
            \textbf{4d $\mathbf{\mathcal{N}\!=\!2}$ $\mathbf{\Omega}$-deformed\\ super Yang-Milles}
            \nodepart{second}\footnotesize\textcolor{red}{SU(2) Nekrasov instanton \\ partition function}
        };
        \node (2dClassCFT) [abstract, rectangle split, rectangle split parts=2,below=of 2dCFT]
        {\small
        \textbf{Class 2d CFT}
            \nodepart{second}\footnotesize\textcolor{red}{classical conformal\\ blocks}
        };
        \node (2dN2SYM) [abstract, rectangle split, rectangle split parts=2,right=of 2dClassCFT]
        {\small
            \textbf{2d $\mathbf{\mathcal{N}\!=\!2}$ $\mathbf{\Omega}$-deformed\\ super Yang-Milles}
            \nodepart{second}\footnotesize\textcolor{red}{SU(2) instanton \\twisted superpotentials}
        };
        \node (QIS) [abstract, rectangle split, rectangle split parts=2,right=of 2dN2SYM]
        {\small
            \textbf{2-particle QIS}
            \nodepart{second}\footnotesize\textcolor{red}{Eigenvalues of some \\ Schr\"{o}dinger operators}
        };
\begin{scope}
\draw[darrow] (2dCFT.east) -- (4dN2SYM.west) 
node[above,desc=center] { \footnotesize\bf \textcolor{blue}{AGT} };
\draw[darrow] (2dClassCFT) -- (2dN2SYM) 
node[above,desc=center] {\footnotesize\bf \textcolor{blue}{Class} } 
node[below,desc=center] {\footnotesize\bf \textcolor{blue}{AGT} };
\draw[darrow] (2dN2SYM) -- (QIS)  node[above,desc=center] {\footnotesize\bf \textcolor{blue}{gauge} }
node[below,desc=center] {\footnotesize \bf \textcolor{blue}{Bethe} };
\draw[sarrow] (2dCFT) -- (2dClassCFT)  
node[left,desc=right] {\footnotesize\bf \textcolor{olive}{classical} }
node[right,desc=left] {\footnotesize\bf \textcolor{olive}{limit} };
\draw[sarrow] (4dN2SYM) -- (2dN2SYM)  
node[left,desc=right] {\footnotesize\bf \textcolor{olive}{Nekrasov\\[-2pt]-Shatashvili} }
node[right,desc=left] { \footnotesize\bf \textcolor{olive}{limit} };
\path[sarrow,draw] (2dClassCFT.south) |- ([yshift=-0.3cm] 2dN2SYM.south) 
node[below] at ([yshift=-0.3cm] 2dN2SYM.south) {\large \textcolor{magenta}{\small Classical limit of null vector decoupling equations} } 
([yshift=-0.3cm] 2dN2SYM.south) -| (QIS.south);
\end{scope}  
\end{tikzpicture}
}
\caption{The triple correspondence in the case of the Virasoro classical conformal blocks
links the latter to SU(2) instanton twisted superpotentials which describe the spectra of some
quantum--mechanical systems. The Bethe/gauge correspondence on the r.h.s.~connects
the SU(N) ${\cal N}=2$ SYM theories with the N--particle quantum integrable systems. 
An extension of the above triple relation to the case ${\rm N}>2$ needs to consider on the l.h.s.~the 
classical limit of the $W_{\rm N}$ symmetry conformal blocks according to the known extension 
\cite{Wyllard:2009hg} of the AGT conjecture
(see e.g.~\cite{Poghossian:2016rzb,Poghosyan:2016lya,Poghosyan:2016mkh}).}
\end{figure}
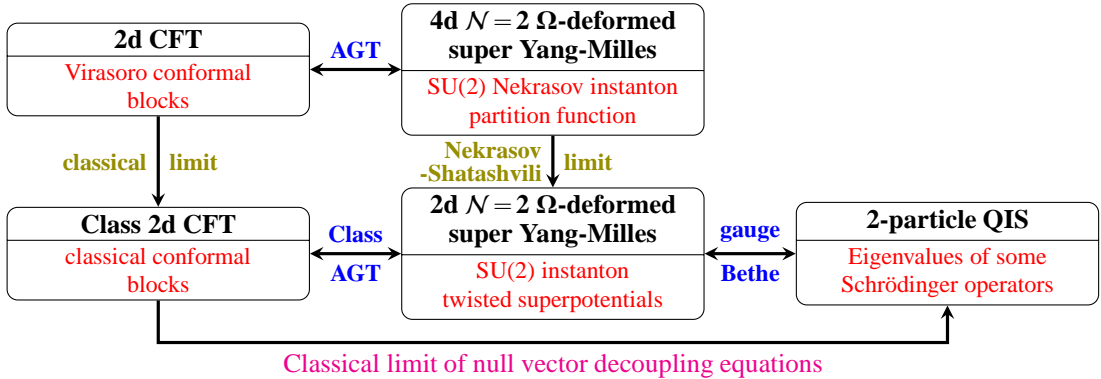

The Mathieu and Whittaker--Hill equations can be studied using conventional, well known
methods such as the Hill determinant or WKB, 
cf.~e.g.~\cite{MagnusWinkler,Whittaker:1996,MuellerKirsten:2006,Kashani-Poor:2015pca}.
On the other hand, it has been observed lately that Schr\"{o}dinger equations with
potentials (\ref{Mathieu})--(\ref{complexQ}) emerge entirely within the framework of
two-dimensional conformal field theory (2dCFT) as the classical limit of the null vector
decoupling (NVD) equations obeyed by certain 3-point degenerate irregular conformal blocks 
\cite{Maruyoshi:2010,Piatek:2014lma,Piatek:2015jva,Rim:2015tsa,Bonelli:2011aa}. 
Moreover, as a manifestation of the correspondence between the ``semiclassical''
2dCFT and the Nekrasov--Shatashvili limit of the $\Omega$-deformed ${\cal N}\!=\!2$ super Yang--Mills theories 
(cf.~Fig.~1) the spectrum of the Mathieu and related operators can be investigated by making use of
tools of the ${\cal N}\!=\!2$ SUSY gauge theories, cf.~e.g.~\cite{Piatek:2014lma,Basar:2015xna}.

In our previous works \cite{Piatek:2014lma,Piatek:2015jva} we have fund in particular that  
the Mathieu eigenvalue can be expressed in terms of the pure gauge classical irregular block and such expression
exactly coincides with the well known weak  coupling expansion of the Mathieu eigenvalue in the case in which the 
auxiliary parameter is the non-integer Floquet exponent. Furthermore, it has been shown that 
the formula for the corresponding eigenfunction obtained from the irregular block reproduces the so-called 
Mathieu exponent from which the non-integer order elliptic cosine and sine functions may be constructed.

In the present paper we continue the line of research initiated in \cite{Piatek:2014lma,Piatek:2015jva}
and study the eigenvalue problem (\ref{HillEq}) for potentials (\ref{WH}) and (\ref{complexQ}) using
methods of 2dCFT. The purpose of this work is to answer the question of what kind of solutions
are possible to be obtained in this way. This knowledge paves the way for studying 
spectra of Schr\"{o}dinger operators with potentials (\ref{WH}) and (\ref{complexQ}) employing
non-perturbative tools of 2dCFT. Precisely, it seems to be possible to connect different regions
of spectra of mentioned operators using duality relations for 
four-point spherical conformal blocks. This is the main motivation
for our research.

The organization of the paper is as follows. In section \ref{section2} we introduce the Gaiotto
vectors (GV) related to the ${\cal N}\!=\!2$ SYM theories with $N_f=0, 1$ flavors and define 
two types of irregular conformal blocks, i.e.:
($a$) the products of GV, and ($b$) the
matrix elements of certain degenerate chiral vertex operators between GV.
Then, we discuss some basic properties of these blocks. In 
particular, we propose a classical asymptotical behavior for irregular blocks of the type ($a$) 
which is inspired by the semiclassical behavior of the physical Liouville field theory correlators
and consistent with the Nekrasov--Shatashvili limit of the corresponding Nekrasov instanton functions. 
Moreover, using this proposal we compute power expansions of 
the {\it classical} $N_f=1, 2$ irregular conformal blocks.\footnote{Cf.~\cite{Rim:2015tsa,Rim:2015aha},
where the classical irregular blocks have been also studied using methods of matrix models.}
Finally, we derive certain null vector decoupling equations obeyed by the $N_f=1, 2$ degenerate irregular blocks 
of the type ($b$).

In section \ref{section3} we derive classical limit of the NVD equations fulfilled by the  
$N_f=1, 2$ degenerate irregular blocks. As a result we get closed expressions for some solutions of 
the eigenvalue problem (\ref{HillEq}) with the {\sf PT}--symmetric complex
potential (\ref{complexQ}) and the Whittaker--Hill potential (\ref{WH}).\footnote{Note that 
an appropriate substitution transforms the Whittaker--Hill 
equation to the so-called equation of Ince, cf. 
[\href{http://dlmf.nist.gov/28.31}{http://dlmf.nist.gov/28.31}].
Hence, our formulae may also be useful to express some solutions of the Ince equation.}
More concretely, we have found that
(i) for each potential (\ref{WH}) and (\ref{complexQ}) the eigenvalue $\lambda$ and the corresponding two
linearly independent solutions of the eq.~(\ref{HillEq}) are given in terms of the classical limit
of irregular blocks;
(ii) for the complex potential (\ref{complexQ}) the spectrum $\lambda$ is indeed real for $h,x\in\mathbb{R}$
and $\nu\in\mathbb{R}\setminus\mathbb{Z}$;
(iii) our fundamental solutions to eqs.~(\ref{HillEq})--(\ref{WH}) and (\ref{HillEq})--(\ref{complexQ}) 
are nothing but the non-integer order Floquet solutions in the weak coupling (small $h$) 
region.\footnote{For the definition and classification of the Floquet solutions
of the Hill equation, see appendix \ref{sec:App_A}.}

In section \ref{section4} we present our conclusions. The problems that are still open and the possible 
extensions of the present work are discussed.

\section{Quantum and classical \texorpdfstring{$N_f=1,2$}{Nf=1,2} irregular blocks}
\label{section2}
As we have mentioned earlier in our previous paper in the sequel
\cite{Piatek:2015jva} the Moore--Seiberg formalism of rational 
conformal field theory can be successfully extended to the 
case of non-rational 2dCFT. Therefore the central role in 
the forthcoming discussion is played by the \emph{chiral vertex operators} (CVO's)
that constitute building blocks for physical fields in the Moore--Seiberg formalism.
CVO's are assumed here to act between highest weight representations of Virasoro
algebra. 

\subsection{Regular and \texorpdfstring{$N_f=1,2$}{Nf=1,2} irregular blocks}
For the sake of definiteness let ${\cal V}_{c, \Delta}^{\,n}$ denote 
the vector space generated by all vectors of the form
\begin{equation}
\label{Basis}
|\,\nu_{\Delta,I}\,\rangle=L_{-I}|\,\nu_{\Delta}\,\rangle := 
L_{-k_{1}}\ldots L_{-k_{j-1}}L_{-k_{j}} |\,\nu_{\Delta}\,\rangle,
\qquad
\underset{i\in\mathbb{N}}{\forall}\quad L_{-i}\in\mathsf{Vir}_{c},
\end{equation}
where
$I=(k_{1},\ldots, k_{j-1}, k_{j})$ is an ordered ($k_{1}\geq \ldots\geq
k_{j}\geq 1$) sequence of positive integers of the length $|I|\equiv k_{1}+\ldots+k_{j}=n$,
and $|\,\nu_{\Delta}\,\rangle$ is the highest weight vector:
\begin{equation}
\label{HW}
L_{0}|\,\nu_{\Delta}\,\rangle=
\Delta|\,\nu_{\Delta}\,\rangle,\qquad\underset{n\in\mathbb{N}}{\forall}
\quad
L_{n}|\,\nu_{\Delta}\,\rangle= 0\,.
\end{equation}
The Verma module of the central charge $c$ and the highest weight $\Delta$ 
is the $\mathbb Z$-graded representation of the Virasoro algebra determined on the space:
$$
{\cal V}_{c,\Delta} = \bigoplus_{n\geq 0}{\cal V}_{c,\Delta}^{\,n}, 
\qquad \dim{\cal V}_{c,\Delta}^{\,n} = p(n)\ ,
$$
where $p(n)$ is the number of partitions of $n$ (with the convention $p(0)=1$).
It is an eigenspace of $L_{0}$ with the eigenvalue $\Delta+n$.
On ${\cal V}_{c,\Delta}^{n}$ there exists the symmetric
bilinear form $\langle\,\cdot\,|\,\cdot\,\rangle$
uniquely defined by the relations
$$
\langle\,\nu_{\Delta}\,|\,\nu_{\Delta}\,\rangle\;=\;1\;\;\;{\rm and}\;\;\;
(L_n)^{\dagger}\;=\;L_{-n}.
$$
The Gram matrix $G_{c,\Delta}$ of the form $\langle\,\cdot\,|\,\cdot\,\rangle$
is block-diagonal in the basis 
$\left\lbrace|\,\nu_{\Delta, I}\,\rangle\right\rbrace$ with blocks
$$
\Big[G_{c,\Delta}^{n}\Big]_{IJ}\;=\;
\langle\, \nu_{\Delta, I}\,|\,\nu_{\Delta, J}\,\rangle
\;=\;\langle\, \nu_{\Delta}\,|(L_{-I})^{\dagger}L_{-J}|\,\nu_{\Delta}\,\rangle.
$$
The Verma module ${\cal V}_{c,\Delta}$ is irreducible if and only if the form 
$\langle\,\cdot\,|\,\cdot\,\rangle$ is non-degenerate.
The criterion for irreducibility is vanishing of the determinant $\det G_{c,\Delta}^{n}$
of the Gram matrix, known as the Kac determinant, given by the formula 
\cite{Kac:1978ge,Feigin:1981st,FF2,Thorn:1984sn,Kac:1987gg}:
\begin{equation}
\label{det} 
\textrm{det}\; G^{\,n}_{c,\Delta} \;=\; C_n
\prod_{1\leq rs \leq n} (\dl - \dl_{rs})^{p(n-rs)} ,
\end{equation}
where  $C_n$ is a constant and $\dl_{rs}$ are the weights form the Kac table
\begin{equation}
\label{def_degenerate_weights}
\Delta_{rs}(c) = \frac{{\sf Q}^2}{4} - \frac{1}{4}\left(rb + \frac{s}{b} \right)^{2} ,
\quad r,s\in\mathbb{N},
\end{equation}
for which the central charge is given by $c =1+6{\sf Q}^2$ with ${\sf Q}=b + b^{-1}$.

The non-zero element $|\,\chi_{rs}\,\rangle\in{\cal V}_{c,\Delta_{rs}(c)}$ of degree 
$n=rs$ is called a null vector if
$\label{zerowydef} L_0\,|\,\chi_{rs}\,\rangle =(\Delta_{rs} +
rs)\,|\,\chi_{rs}\,\rangle$,
and
$L_{k}\,|\,\chi_{rs}\,\rangle = 0$, $\forall\,k>0$.
Hence, $|\,\chi_{rs}\,\rangle$ is the highest weight state which generates its own Verma
module ${\cal V}_{c,\Delta_{rs}(c)+rs}$,
which is a submodule of ${\cal V}_{c,\Delta_{rs}(c)}$. One can prove that
each submodule of the Verma module ${\cal V}_{c,\Delta_{rs}(c)}$ is generated by
a null vector. Then,  the module ${\cal V}_{c,\Delta_{rs}(c)}$ is irreducible if and only if
it does not contain null vectors with positive degree.

For non-degenerate values of $\Delta$, i.e.~for $\Delta\neq\Delta_{rs}(c)$, there
exists the basis $\lbrace|\,\nu^{t}_{\Delta, I}\,\rangle\rbrace_{I\vdash n}$ in ${\cal V}_{c,\Delta}^{n}$
whose elements are defined by the relation
$
\langle\,\nu^{t}_{\Delta, I}\,|\,\nu_{\Delta, J}\,\rangle = \delta_{IJ}
$
for all
$|\,\nu_{\Delta, J}\,\rangle \in{\cal V}_{c,\Delta}^{n}$.
The basis vectors $|\,\nu^{t}_{\Delta,I}\,\rangle$ have
the following representation in the standard basis
$$
|\,\nu^{t}_{\Delta, I}\,\rangle :=\sum_{J\vdash n}
\Big[ G_{c,\Delta}^n\Big]^{IJ} |\,\nu_{\Delta, J}\,\rangle,
\qquad \sum_{K\vdash n} \Big[ G_{c,\Delta}^n\Big]^{IK}\Big[ G_{c,\Delta}^n\Big]_{K J} = \d^{I}_{J}.
$$

The chiral vertex operator is the linear map
$
V{^{\Delta_3}_\infty}{^{\Delta_2}_{\:z}}{^{\Delta_1}_{\;0}} :
{\cal V}_{\Delta_2} \otimes {\cal V}_{\Delta_1}
\to
{\cal V}_{\Delta_3}
$
such that for all $|\,\xi_2\,\rangle \in {\cal V}_{\Delta_2}$ the operator
$$
V(\xi_2 | z) \equiv
V{^{\Delta_3}_\infty}{^{\Delta_2}_{\:z}}{^{\Delta_1}_{\;0}}
(|\,\xi_2\,\rangle\otimes\,\cdot\,):{\cal V}_{\Delta_1}
\to
{\cal V}_{\Delta_3}
$$
satisfies the following conditions
\begin{eqnarray}
\label{CVO}
\left[L_n , V\!\left(\nu_{2}|z\right)\right] &=& z^{n}\left(z
\frac{\partial}{\partial z} + (n+1)\Delta_2
\right) V\!\!\left(\nu_{2}|z\right)\,,\;\;\;\;\;\;\;\;n\in\mathbb{Z}
\\
\label{CVO2}
V\!\!\left(L_{-1}\xi_{2}|z\right) &=& \frac{\partial}{\partial z}V\!\!\left(\xi_{2}|z\right),
\\
\label{CVO3}
V\!\!\left(L_{n}\xi_{2}|z\right) &=& \sum\limits_{k=0}^{n+1}
\left(\,_{\;\;k}^{n+1}\right)
(-z)^{k}\left[L_{n-k}, V\!\!\left(\xi_{2}|z\right)\right]\,,
\;\;\;\;\;\;\;\;n>-1,
\\
V\!\!\left(L_{-n}\xi_{2}|z\right) &=&\sum\limits_{k=0}^{\infty}
\left(\,_{\;\;n-2}^{n-2+k}\right)
z^k\,L_{-n-k}\,V\!\!\left(\xi_{2}|z\right)
\nonumber
\\
\label{CVO5}
&+& (-1)^{n}\sum\limits_{k=0}^{\infty}
\left(\,_{\;\;n-2}^{n-2+k}\right)
z^{-n+1-k}\,\,V\!\!\left(\xi_{2}|z\right)\,L_{k-1},
\;\;\;\;\;n>1.
\end{eqnarray}
The commutation relation (\ref{CVO}) defines the primary vertex operator
corresponding to the highest weight state 
$|\,\nu_2\,\rangle\in{\cal V}_{\Delta_2}$.
The matrix element of the primary CVO between basis states in $\mathcal{V}^{|I|,|J|}_{\dl_{a,c}}$
fulfills the following relation
\begin{equation}
\label{CVO_normalization}
\langle\,\nu_{\Delta_a,I}\,|
V\!\left(\nu_{\Delta_{b}} |z\right)
|\,\nu_{\dl_{c},J}\,\rangle
= z^{\Delta_a -\Delta_{b}-\Delta_c +|I| - |J|}
\bra{\n_{\dl_{a}}} V\!\left(\nu_{\Delta_{b}} |1\right) \ket{\n_{\dl_{c}}}  .
\end{equation}
In what follows we assume the normalization 
$\bra{\n_{\dl_{a}}} V\!\left(\nu_{\Delta_{b}} |1\right) \ket{\n_{\dl_{c}}} = 1$.

In analogy to the Heissenberg algebra it is possible to form coherent states 
for generators of the Virasoro algebra. Indeed, as it was first shown by Gaiotto \cite{Gaiotto:2009}
and in refs. \cite{Marshakov:2009,Bonelli:2011aa,Felinska:2011tn,Gaiotto:2012sf} the Virasoro
generators for each $n>0$ in the highest weight condition in eq. \eqref{HW} 
may be treated as ``annihilation'' operators. 
Hence, there is a vector on which a certain finite set of positive indexed generators of Virasoro
algebra act diagonally termed the irregular vector. The Virasoro algebra each 
positive indexed generator obeys induces corresponding algebra of differential operators in the space
of parameters labeling the irregular vector \cite{Gaiotto:2012sf}. In what follows we are concerned
with only two simplest cases that due to AGT correspond to gauge theories with $N_{f}=0$ and $N_{f}=1$, 
where there are at most two parameters labeling the irregular vector.\footnote{In the 
present paper we adopt the following nomenclature:
``zero flavor'', ``single flavor'' and so on for Gaiotto's states and then for 
irregular blocks as these objects are related to corresponding quantities in 
the ${\cal N}=2$ super-Yang--Mills theories with $N_f=0,1,\ldots$ 
--- zero, one and more flavors.}

Let us recall that the Virasoro irregular vector that is a ``coherent state'' of $L_{1}$ to which we
further refer as to the ``zero flavor'' or $N_{f}=0$ state fulfills the following conditions:
\begin{subequations}
\begin{displaymath}
L_{0}|\Delta,\Lambda^2\rangle =
\left(\Delta\!+\!\frac{\Lambda}{2}\frac{\partial}{\partial\Lambda}\right)|\Delta,\Lambda^2\rangle,
\quad
L_{1}|\Delta,\Lambda^2\rangle = \Lambda^2 |\Delta,\Lambda^2\rangle,
\quad
L_{n}|\Delta,\Lambda^2\rangle = 0 \;\;\forall\;n\geq 2.
\end{displaymath}
Making use of the above algebra it can be
developed in ${\cal V}_{c,\Delta}$  as follows\footnote{The symbol $\mathbb{Y}$
denotes the set of all partitions or equivalently the Young diagrams of all 
natural numbers that in the mathematical literature is termed the Young lattice. }
\begin{equation}
\label{GaiottoNf=0}
\ket{\dl,\La^{2}} 
= \sum_{I\in\mathbb{Y}} \ket{\n_{\dl,I}}\bracket{\n_{\dl,I}^{t}}{\dl,\La^{2}}
= \sum\limits_{n=0}^{\infty}\Lambda^{2n}\sum\limits_{I\vdash n}
\Big[G^{n}_{c,\Delta}\Big]^{(1^{n}) I}|\,\nu_{\Delta,I}\,\rangle\;,
\tag{\theequation G1}
\end{equation} 
where we used the projector (identity operator in ${\cal V}_{c,\Delta}$):
$$
\mathbb{P}_{\Delta}
= \sum_{I\in\mathbb{Y}} \ket{\n_{\dl,I}}\otimes\bra{\n_{\dl,I}^{t}}
= \sum_{n\geq 0}
\sum_{I,J\vdash n}\sbr{G_{c,\dl}^{n}}^{I J} \ket{\n_{\dl,I}}\otimes\bra{\n_{\dl,J}}\; ,
$$
and a convenient normalization $\bracket{\n_{\dl}}{\dl,\La^{2}}=1$ .
\end{subequations}
The irregular vector to which we further refer as to the ``single flavor'' or $N_{f}=1$ state 
fulfills the following defining conditions 
\begin{align*}
L_{0}|\,\Delta,\Lambda, m\,\rangle&
=\left(\Delta+\Lambda\frac{\partial}{\partial\Lambda}\right)|\,\Delta,\Lambda, m\,\rangle,
&
L_{1}|\,\Delta,\Lambda, m\,\rangle & = m\Lambda |\,\Delta,\Lambda,m\,\rangle,
\\
L_{2}|\,\Delta,\Lambda, m\,\rangle &=\Lambda^2 |\,\Delta,\Lambda,m\,\rangle,
&
L_{n}|\,\Delta,\Lambda, m\,\rangle &= 0\quad \forall\;n\geq 3.
\end{align*}
Proceeding analogously as in $N_{f}=0$ case we obtain
\begin{equation}
\label{GaiottoNf=1}
|\,\Delta,\Lambda, m\,\rangle\;
= \sum\limits_{n=0}^{\infty}\Lambda^{n}\sum\limits_{p=0}^{[\frac{n}{2}]} m^{n-2p}
\sum\limits_{I\vdash n}\Big[G^{n}_{c,\Delta}\Big]^{(1^{n-2p}\,2^p) I}
|\,\nu^n_{\Delta,I}\,\rangle\;.
\tag{\theequation G2}
\end{equation}
The quantum irregular conformal blocks are defined as inner products of the 
irregular vectors \cite{Gaiotto:2009,Marshakov:2009}:\footnote{In fact, there is much more 
Gaiotto's states than just these two written in eqs.~(\ref{GaiottoNf=0}) 
and (\ref{GaiottoNf=1}), see for instance 
\cite{Bonelli:2011aa,Felinska:2011tn,Gaiotto:2012sf}.
In the present paper we confine ourselves to study irregular 
blocks being inner products of (\ref{GaiottoNf=0}), (\ref{GaiottoNf=1})
or matrix elements between these two states. Possible extensions of the
present work taking into account existence of the other Gaiotto states are
discussed in conclusions.}
\begin{eqnarray}
\label{FNf1}
{\cal F}_{c,\Delta}(\Lambda, m) &=&
\langle\,\Delta, \tfrac{1}{2}\Lambda, 2m\,|\,\Delta,\Lambda^2\,\rangle
\\
&=&
\sum\limits_{n=0}^{\infty}\left(\tfrac{1}{2}\Lambda^3\right)^n
\sum\limits_{p=0}^{[\frac{n}{2}]}\left(2m\right)^{n-2p}
\Big[G^{n}_{c,\Delta}\Big]^{(1^{n-2p}\,2^p)(1^n)},\nonumber
\\[8pt]
\label{FNf2}
{\cal F}_{c,\Delta}(\Lambda, m_1, m_2) &=&
\langle\,\Delta, \tfrac{1}{2}\Lambda, 2m_1\,|\,\Delta,\tfrac{1}{2}\Lambda, 2m_2\,\rangle
\\
&=&
\sum\limits_{n=0}^{\infty}\left(\frac{\Lambda}{2}\right)^{2n}\sum\limits_{p,p'=0}^{[\frac{n}{2}]}
(2 m_1)^{n-2p}\Big[G^{n}_{c,\Delta}\Big]^{(1^{n-2p}\,2^p)\,(1^{n-2p}\,2^p)}
(2 m_2)^{n-2p'}.\nonumber
\end{eqnarray}
Irregular blocks (\ref{FNf1})-(\ref{FNf2})
can be recovered from conformal blocks
on the torus and on the sphere in a properly defined decoupling limits of 
external conformal weights \cite{Marshakov:2009,Alba:2009fp}. 
To see this let $C_{g,n}$ denote the Riemann surface with genus $g$ and $n$ punctures.
Let $x$ be the modular parameter of the 4-punctured Riemann sphere 
then the $s$-channel conformal block on $C_{0,4}$ is defined as the 
following formal $x$-expansion:
\begin{subequations}
\begin{equation}
\label{four-pointblock}
{\cal F}_{c,\Delta}\!\left[_{\Delta_{1}\;\Delta_{4}}^{\Delta_{2}\;\Delta_{3}}\right]\!(x)=
x^{\Delta-\Delta_{3}-\Delta_{4}}\left( 1 +
\sum_{n=1}^\infty x^{\,n}
{\cal F}^{\,n}_{c,\Delta}\!\left[_{\Delta_{1}\;\Delta_{4}}^{\Delta_{2}\;\Delta_{3}}\right] \right),
\end{equation}
where the coefficients of the conformal block are defined as 
\begin{equation}
\label{4pBC}
{\cal F}^{\,n}_{c,\Delta}\!\left[_{\Delta_{1}\;\Delta_{4}}^{\Delta_{2}\;\Delta_{3}}\right]
= \sum\limits_{I,J\vdash n}
\bra{\n_{\dl_{1}}}V_{\dl_{2}}(1)\ket{\n_{\dl,I}^{n}}
\Big[ G_{c,\Delta}^{n}\Big]^{IJ}
\bra{\n_{\dl,J}^{n}}V_{\dl_{3}}(1)\ket{\n_{\dl_{4}}} .
\end{equation} 
\end{subequations}
Now, employing a suitable AGT inspired parametrization 
of the external weights $\Delta_i$ and the central charge $c$, i.e.:
\begin{gather*}
\Delta_i = \frac{\alpha_i(\epsilon - \alpha_i)}{\epsilon_1\epsilon_2}\,,
\quad
c = 1 + 6\frac{\epsilon^2}{\epsilon_1\epsilon_2},
\qquad
\epsilon = \epsilon_1 + \epsilon_2\; ,
\\
\alpha_1 = \tfrac{1}{2}\left(\epsilon+\mu_1-\mu_2\right), 
\quad 
\alpha_2 = \tfrac{1}{2}\left(\mu_1+\mu_2\right),
\quad 
\alpha_3 = \tfrac{1}{2}\left(\mu_3 + \mu_4\right), 
\quad 
\alpha_4 = \tfrac{1}{2}\left(\epsilon+\mu_3-\mu_4\right),
\end{gather*}
and introducing the dimensionless expansion parameters:
$\Lambda=\hat\Lambda/(-\epsilon_1\epsilon_2)^{\frac{1}{2}}$ and
$m_i=\hat m_i/(-\epsilon_1\epsilon_2)^{\frac{1}{2}}$, 
where
$\hat m=\mu_1-\frac{1}{2}\epsilon$,
$\hat m_{1,4}=\mu_{1,4}-\frac{1}{2}\epsilon$,
one obtains \cite{Marshakov:2009,Alba:2009fp}:
\begin{eqnarray*}
x^{\Delta_3+\Delta_4-\Delta}
{\cal F}_{c,\Delta}\!\left[_{\Delta_{1}\;\Delta_{4}}^{\Delta_{2}\;\Delta_{3}}\right]\!(x)
&\xrightarrow[x\mu_2\mu_3\mu_4=\hat\Lambda^3]{\mu_2,\mu_3,\mu_4\,\to\,\infty}&
{\cal F}_{c,\Delta}(\Lambda, m),
\\[8pt]
x^{\Delta_3+\Delta_4-\Delta}
{\cal F}_{c,\Delta}\!\left[_{\Delta_{1}\;\Delta_{4}}^{\Delta_{2}\;\Delta_{3}}\right]\!(x)
&\xrightarrow[x\mu_2\mu_3=\hat\Lambda^2]{\mu_2,\mu_3\,\to\,\infty}&
{\cal F}_{c,\Delta}(\Lambda, m_1, m_4).
\end{eqnarray*}
The above conformal blocks may be related by means of AGT correspondence to their
gauge theoretic counterparts, that is $SU(2)$ Nekrasov's instanton partition functions with $N_f=1,2$ flavors 
\cite{Gaiotto:2009,Hadasz:2010xp,Tan:2013tq,Maulik:2012wi}:
\begin{eqnarray}
\label{AGT1}
{\cal F}_{c,\Delta}(\Lambda, m) &=&
{\cal Z}_{\rm inst}^{SU(2), N_f=1}(\hat\Lambda, a, \hat m, \epsilon_1, \epsilon_2),
\\[8pt]
\label{AGT2}
{\cal F}_{c,\Delta}(\Lambda, m_1, m_2) &=&
{\cal Z}_{\rm inst}^{SU(2), N_f=2}(\hat\Lambda, a, \hat m_1, \hat m_2, \epsilon_1, \epsilon_2).
\end{eqnarray}
The relations~(\ref{AGT1})-(\ref{AGT2}), which are understood as equalities between the coefficients of the
expansions of both sides, hold for
\begin{equation}
\label{para1}
\Lambda=\frac{\hat\Lambda}{\sqrt{-\epsilon_1\epsilon_2}},
\qquad
m_i=\frac{\hat m_i}{\sqrt{-\epsilon_1\epsilon_2}},
\qquad
\Delta = \frac{\epsilon^2 - 4a^2}{4\epsilon_1\epsilon_2},
\qquad
c=1 + 6\frac{\epsilon^2}{\epsilon_1\epsilon_2}
\equiv 1+6{\sf Q}^2
\end{equation}
where
\begin{equation}
\label{para2}
{\sf Q}\;=\;b+\frac{1}{b}\;\equiv\;\sqrt{\frac{\epsilon_2}{\epsilon_1}} + \sqrt{\frac{\epsilon_1}{\epsilon_2}},
\qquad  b=\sqrt{\frac{\epsilon_2}{\epsilon_1}}.
\end{equation}

The study of the Nekrasov partition functions ${\cal Z}_{\rm Nekrasov}={\cal Z}_{\rm pert}{\cal Z}_{\rm inst}$
in the limit $\epsilon_2\to 0$ has revealed that it behaves exponentially \cite{NS:2009}. 
In particular, for the instanton part of the partition function we have
\begin{equation}
\label{asymZ}
\mathcal{Z}_{\rm inst}(\,\cdot\,,\epsilon_1,\epsilon_2)\;\stackrel{\epsilon_2\to 0}{\sim}\;
\exp\left\lbrace \frac{1}{\epsilon_2}\,\mathcal{W}_{\rm inst}(\,\cdot\,,\epsilon_1)\right\rbrace.
\end{equation}
Therefore, in view of eqs.~(\ref{AGT1})-(\ref{AGT2}), and relation between Liouville and deformation
parameters in eq. \eqref{para2} from the Nekrasov--Shatashvili 
limit in eq. (\ref{asymZ}) one can expect the exponential behavior 
of irregular blocks in the limit $b\to 0$, i.e.:
\begin{eqnarray}
\label{ClIrrblock1}
{\cal F}_{1+6Q^2,\Delta}(\Lambda, m)
&\stackrel{b\to 0}{\sim}&
\exp\left\lbrace\frac{1}{b^2}f_{\delta}^{\bf 1}\!\left(\hat\Lambda/\epsilon_1, 
\hat m/\epsilon_1\right)\right\rbrace,
\\[8pt]
\label{ClIrrblock2}
{\cal F}_{1+6Q^2,\Delta}(\Lambda, m_1, m_2)
&\stackrel{b\to 0}{\sim}&
\exp\left\lbrace\frac{1}{b^2}f_{\delta}^{\bf 2}\!\left(\hat\Lambda/\epsilon_1, 
\hat m_1/\epsilon_1, \hat m_2/\epsilon_1\right)\right\rbrace,
\end{eqnarray}
where $\Delta=\frac{1}{b^2}\delta$, $\delta={\cal O}(b^0)$.
The classical behaviors (\ref{ClIrrblock1})-(\ref{ClIrrblock2}) are 
very nontrivial statements concerning quantum irregular blocks.
Although there is no proof of eqs.~(\ref{ClIrrblock1})-(\ref{ClIrrblock2}) 
the existence of the {\it classical irregular blocks} $f^{\bf 1}$, $f^{\bf 2}$
can be verified through direct calculation order by order in $\hat\La/\e_{1}$. 
Using power expansions of quantum irregular 
blocks (\ref{FNf1})-(\ref{FNf2})
and eqs.~(\ref{ClIrrblock1})-(\ref{ClIrrblock2}) up to $n=3$ one finds
\begin{itemize}
\item[$(i)$] the classical irregular block with ``single flavor'' $N_f=1$:
\begin{subequations}
\begin{eqnarray}
\label{ClIrrexp2}
f_{\delta}^{\bf 1}\!\left(\hat\Lambda/\epsilon_1, \hat m/\epsilon_1\right) &=&
\lim\limits_{b\to 0} b^2 \log {\cal F}_{1+6Q^2, \frac{1}{b^2}\delta}
\!\left(\hat\Lambda/(\epsilon_1 b), \hat m/(\epsilon_1 b)\right)\nonumber
\\
&=&
\sum\limits_{n=1}^{\infty}
\left(\hat\Lambda/\epsilon_1\right)^{\!3n}
\!f_{\delta}^{{\bf 1},n}\!\left(\frac{\hat m}{\epsilon_1}\right),
\end{eqnarray}
where the coefficients read as follows
\begin{equation}
\label{f1coeffs}
\begin{aligned}
f^{\mathbf{1},1}_{\delta}\!\left(\frac{\hat{m}}{\epsilon_1}\right) &= \frac{1}{2\delta }\frac{\hat{m}}{\epsilon _1} ,
\\
f^{\mathbf{1},2}_{\delta}\!\left(\frac{\hat{m}}{\epsilon_1}\right) 
&= \frac{5 \delta -3}{16 \delta ^3 (4\delta +3)}\left(\frac{\hat{m}}{\epsilon_1}\right)^2
-\frac{3}{16 \delta  (4\delta +3)} ,
\\
f^{\mathbf{1},3}_{\delta}\!\left(\frac{\hat{m}}{\epsilon_1}\right) &
= \frac{\delta(9 \delta -19)+6}{48 \delta ^5 (\delta+2) (4 \delta +3)}\left(\frac{\hat{m}}{\epsilon_1}\right)^3
+\frac{6-7\delta }{48\delta^3 (\delta +2) (4 \delta +3)}\frac{\hat{m}}{\epsilon _1} ;
\end{aligned}
\end{equation}
\end{subequations}

\item[$(ii)$] the classical irregular block with ``two flavors'' $N_f=2$:
\begin{subequations}
\begin{eqnarray}
\label{ClIrrexp3}
f_{\delta}^{\bf 2}\!\left(\hat\Lambda/\epsilon_1, \hat m_1/\epsilon_1, \hat m_2/\epsilon_1\right) 
&=&
\lim\limits_{b\to 0} b^2 \log {\cal F}_{1+6Q^2, \frac{1}{b^2}\delta}
\!\left(\hat\Lambda/(\epsilon_1 b), \hat m_2/(\epsilon_1 b), \hat m_1/(\epsilon_1 b)\right)\nonumber
\\
&=&
\sum\limits_{n=1}^{\infty}
\left(\hat\Lambda/\epsilon_1\right)^{\!2n}
\!f_{\delta}^{{\bf 2},n}\!\left(\frac{\hat m_1}{\epsilon_1},\frac{\hat m_2}{\epsilon_1}\right),
\end{eqnarray}
where the coefficients are of the form
\begin{equation}
\label{f2coeffs}
\begin{aligned}
f^{\mathbf{2},1}_{\delta}\left(\frac{\hat{m}_1}{\epsilon_1},\frac{\hat{m}_2}{\epsilon_1}\right) 
&= \frac{1}{2 \delta } \frac{\hat{m}_1}{\epsilon _1} \frac{\hat{m}_2}{\epsilon _1},
\\
f^{\mathbf{2},2}_{\delta}\left(\frac{\hat{m}_1}{\epsilon_1},\frac{\hat{m}_2}{\epsilon_1}\right) 
&= \frac{\delta ^2 \left(\delta -3 \left(\frac{\hat{m}_2}{\epsilon _1}\right){}^2\right)
+\left(\frac{\hat{m}_1}{\epsilon _1}\right){}^2
   \left((5 \delta -3) \left(\frac{\hat{m}_2}{\epsilon _1}\right){}^2-3 \delta ^2\right)}{16 \delta ^3 (4 \delta +3)} ,
\\
f^{\mathbf{2},3}_{\delta}\left(\frac{\hat{m}_1}{\epsilon_1},\frac{\hat{m}_2}{\epsilon_1}\right) 
&
= \frac{1}{48 \delta ^5 (\delta +2) (4 \delta +3)}
\frac{\hat{m}_1}{\epsilon _1} \frac{\hat{m}_2}{\epsilon _1} 
\left[(6-7 \delta ) \delta ^2 \left(\frac{\hat{m}_2}{\epsilon
   _1}\right){}^2\right.
\\
&
\left. +\left(\frac{\hat{m}_1}{\epsilon _1}\right){}^2 \left((\delta  (9 \delta -19)+6) \left(\frac{\hat{m}_2}{\epsilon
   _1}\right){}^2+(6-7 \delta ) \delta ^2\right)+5 \delta ^4\right] .
\end{aligned}
\end{equation}
\end{subequations}
\end{itemize}
As another yet consistency check of our approach
let us observe that combining~(\ref{AGT1})-(\ref{para2}) and 
(\ref{ClIrrblock1})-(\ref{ClIrrblock2}) one gets an identification between 
classical irregular blocks and $SU(2)$ $N_f=1,2$ {\it effective twisted superpotentials}:
\begin{equation}\label{ClassAGT}
f_{\frac{1}{4}-\frac{a^2}{\epsilon_{1}^2}}^{\bf 1,\,2}\!\left(\hat\Lambda/\epsilon_1,\,\cdot\,\right) =
\frac{1}{\epsilon_1}\,\mathcal{W}_{\rm inst}^{SU(2),\,N_f=1,2}\!
\left(\hat\Lambda, a,\,\cdot\,,\epsilon_1\right).
\end{equation}
Note that the {\it classical conformal weight} $\delta$ in the eq.~(\ref{ClassAGT}) above is expressed 
in terms of the gauge theory parameters $a$, $\epsilon_1$. Indeed, one finds  
$$
\delta = \lim\limits_{b\to 0}b^2\Delta 
= \lim\limits_{\epsilon_2\to 0}\frac{\epsilon_2}{\epsilon_1}\Delta
= \frac{1}{4}-\frac{a^2}{\epsilon_{1}^2}.
$$
By means of the expansions (\ref{ClIrrexp2})-(\ref{ClIrrexp3})
and theirs analogues for the twisted superpotentials obtained independently from the instanton
partition functions one can confirm the identities (\ref{ClassAGT}) up to desired order. 

\subsection{Null vector decoupling equations}
In this subsection we shall derive partial differential equations 
obeyed by the {\it degenerate irregular blocks}, cf.~\cite{Maruyoshi:2010}. 
The latter we define as matrix elements of the degenerate chiral vertex operator,\footnote{
Calculations presented in this subsection hold also for $V_{\Delta_{12}}(z)$.}
$V_{+}(z)\!=\!V(|\nu_{\Delta_+}\rangle|z)$, where
$\Delta_+ \!\equiv\! \Delta_{21}= -\frac{3}{4}b^2-\frac{1}{2}$,
between Gaitto's states \eqref{GaiottoNf=0}-\eqref{GaiottoNf=1}:
\addtocounter{equation}{1}
\begin{align}
\label{P1}
\Psi^{{\bf 1}}(z;\Lambda,m) &:=
\langle\,\Delta', \tfrac{1}{2}\Lambda, 2m\,|V_{+}(z)|\,\tilde\Delta, \Lambda^2\,\rangle,
\tag{\theequation $\Psi$1}
\\
\label{P2}
\Psi^{{\bf 2}}(z;\Lambda,m_1, m_2) &:=
\langle\,\Delta', \tfrac{1}{2}\Lambda, 2m_1\,|
V_{+}(z)|\,\tilde\Delta, \tfrac{1}{2}\Lambda, 2m_2\,\rangle .
\tag{\theequation $\Psi$2}
\end{align}

One of the tools that allows to derive the so-called null vector decoupling (NVD) equations
is the following 
\begin{quotation}
\noindent
\textbf{Theorem}(Feigin--Fuchs~\cite{FF2}, cf.~\cite{Teschner:2001rv})
Let $i,j,k\in\left\lbrace 1,2,3\right\rbrace$ be chosen such that $j\neq i$, $k\neq i$, $j\neq k$.
Let us assume that
\begin{enumerate}
\item 
$\Delta_i=\Delta_{rs}\equiv\tfrac{1}{4}{\sf Q}^2-\tfrac{1}{4}\left(rb+sb^{-1}\right)^2$,
$r,s\in\mathbb{N}$;
\item
the vector $\ket{\xi_i}$ lies in the singular submodule generated by the null vector $\ket{\chi_{rs}}$,
i.e.: $\ket{\xi_i}\in{\cal V}_{c,\Delta_{rs}(c)+rs}\subset{\cal V}_{c,\Delta_{rs}(c)}$.
\end{enumerate}
Then, $\bra{\xi_3}V(\xi_2 | z_2)\ket{\xi_1}=0$ if and only if 
\begin{equation}
\label{Dbeta}
\Delta_j = \Delta_{\beta_j} \equiv \frac{{\sf Q}^2}{4}-\frac{1}{4}\beta_{j}^{2}
\;\;\;\;\;{\rm and}\;\;\;\;\;
\Delta_k = \Delta_{\beta_k} \equiv \frac{{\sf Q}^2}{4}-\frac{1}{4}\beta_{k}^{2}
\end{equation}
satisfy the fusion rules $\beta_j - \beta_k = pb+qb^{-1}$,
where $p\in\lbrace 1-r, 3-r, \ldots, r-1\rbrace$ and $q\in\lbrace 1-s, 3-s, \ldots, s-1\rbrace$.
\end{quotation}

We will apply the Feigin--Fuchs Theorem in the case when $\Delta_2=\Delta_{+}\equiv\Delta_{21}$.
Therefore, we have to assume that the weights $\Delta_1=\tilde\Delta$ 
and $\Delta_3=\Delta'$ of {\it in} and {\it out} states are related by the fusion rule I or II:
\begin{eqnarray}
\label{fv1a}
{\rm I}: && \tilde\Delta=\Delta_{\beta_1}, \;\;\;\; \Delta'=\Delta_{\beta_3}=\Delta_{\beta_1+b}
\;\;\;\;\Leftrightarrow\;\;\;\;\beta_3=\beta_1+b,
\\
\label{fv1b}
{\rm II}: && \tilde\Delta=\Delta_{\beta_1}, \;\;\;\; \Delta'=\Delta_{\beta_3}=\Delta_{\beta_1-b}
\;\;\;\;\Leftrightarrow\;\;\;\;\beta_3=\beta_1-b.
\end{eqnarray}
In our calculation we will use a little bit modified (compering to (\ref{Dbeta})) 
parametrization of conformal weights, namely
\begin{equation}\label{Delta_sigma}
\Delta(\sigma)\equiv\frac{{\sf Q}^2}{4}-\sigma^2,
\end{equation}
in which the fusion rules assumed above reads as follows
\begin{eqnarray}
\label{fv2a}
{\rm I}: && 
\tilde\Delta=\Delta\!\left(\sigma-\tfrac{b}{4}\right), 
\;\;\;\;\;\;\;\;
\Delta'=\Delta\!\left(\sigma+\tfrac{b}{4}\right),
\\
\label{fv2b}
{\rm II}: && 
\tilde\Delta=\Delta\!\left(\sigma+\tfrac{b}{4}\right), 
\;\;\;\;\;\;\;\;
\Delta'=\Delta\!\left(\sigma-\tfrac{b}{4}\right).
\end{eqnarray}
The matrix elements such as (\ref{P1}) or (\ref{P2}), i.e.~with the conformal weights fulfilling 
(\ref{fv1a}), (\ref{fv1b}) or (\ref{fv2a}), (\ref{fv2b}) 
will be denoted by $\Psi^{\bf i}_{\rm I}$, $\Psi^{\bf i}_{\rm II}$.

Hence, by virtue of the Feigin--Fuchs Theorem we have four equations for 
$N_{f}=1,2$ written in a concise form as
\begin{eqnarray}
\label{NullFMatrixEl_Nf12}
\bra{\dl',\tfrac{1}{2}\La, 2m}\chi_{+}(z)\ket{\tilde\dl,\; \cdot \;} = 0\; ,
\end{eqnarray}
where
\begin{list}{}{\itemindent=2mm \parsep=0mm \itemsep=0mm \topsep=0mm}
\item[$(a)$] $\chi_{+}(z)$ is the null vertex operator,
\begin{equation}
\label{nullfield}
\chi_{+}(z)=\left(\widehat{L}_{-2}(z) -
\frac{3}{2(2\Delta_{+} +1)}
\,\widehat{L}_{-1}^{\,2}(z)\right)V_{+}(z)
\equiv
V\!\left(\left(L_{-2}+\tfrac{1}{b^2}L_{-1}^{2}\right)\!|\,\nu_{\Delta_+}\,\rangle\,|\,z\,\right) ,
\end{equation}
which corresponds to the null vector 
$$
|\,\chi_{+}\,\rangle =
\chi_{+}(0)|\,0\,\rangle =
\left(L_{-2}+\frac{1}{b^2}L_{-1}^{2}\right)\!|\,\nu_{\Delta_+}\,\rangle ,
$$
from the second level of the Verma module ${\cal V}_{\Delta_+}$;
\item[$(b)$] 
the dot stands for the set of parameters $\La^{2}$ and $\tfrac{1}{2}\La, 2m$
for ``zero flavor'' and ``single flavor'' irregular vector, respectively;
\item[$(c)$] 
the conformal weights obey (\ref{fv2a}) or (\ref{fv2b}).
\end{list}

In order to convert eqs.~\eqref{NullFMatrixEl_Nf12}
to PDE's obeyed by the degenerate irregular blocks
$\Psi^{\bf i}_{\iota}$, ${\bf i} = {\bf 1},{\bf 2}$, 
$\iota={\rm I},{\rm II}$
one needs to employ the following Ward identities
\begin{equation}
\label{CWI}
\bra{\dl',\tfrac{1}{2}\La, 2m}T(w)V_{+}(z)\ket{\tilde\dl,\;\cdot\;}
=\left[
\frac{z}{w(w-z)}\frac{\partial}{\partial z}
+\frac{\Delta_{+}}{(w-z)^2} + U^{\bf i}_{\iota}
\right]\Psi^{\bf i}_{\iota},
\end{equation} 
where\footnote{Here, $U^{\bf i}_{\iota}$, ${\bf i} = {\bf 1},{\bf 2}$
depend on $\iota={\rm I},{\rm II}$ via terms $\Delta'+2\tilde\Delta-\Delta_{+}$
and $\Delta'+\tilde\Delta-\Delta_{+}$.}
\begin{align}
\label{potential_Nf1}
U^{\bf 1}_{\iota} &= 
\tfrac{1}{4}\Lambda^2+\frac{m\Lambda}{w}+\frac{\Lambda^2}{w^3}
+\frac{1}{3w^2}\left(\Lambda\frac{\partial}{\partial\Lambda}
+\Delta'+2\tilde\Delta-\Delta_{+}-z\frac{\partial}{\partial z}\right)\;,
\tag{\theequation U1}
\\
\label{potential_Nf2}
U^{\bf 2}_{\iota} &= \tfrac{1}{4}\Lambda^2
+\frac{m_1 \Lambda}{w}+\frac{m_2 \Lambda}{w^3}+\frac{\frac{1}{4}\Lambda^2}{w^4}+
\frac{1}{2w^2}\left(\Lambda\frac{\partial}{\partial\Lambda}
+\Delta'+\tilde\Delta-\Delta_{+}-z\frac{\partial}{\partial z}\right)\;.
\tag{\theequation U2}
\end{align}
From eq.~\eqref{CWI} and using the formula \cite{Belavin:1984vu}:
$$
\widehat{L}_{-k}(z)\;=\;\frac{1}{2\pi i}\oint\limits_{C_z}dw (w-z)^{1-k}\,T(w),
$$
one finds that 
\begin{equation}
\label{L2}
\bra{\dl',\tfrac{1}{2}\La, 2m}\hat L_{-2}(z)V_{+}(z)\ket{\tilde\dl,\;\cdot\;}
= \sbr{
-\frac{1}{z}\frac{\partial}{\partial z} + U^{\bf i}_{\iota}\big|_{w\to z}
}\Psi^{\bf i}_{\iota},
\end{equation} 
where $U^{\bf i}_{\iota}$ are given in eqs.~\eqref{potential_Nf1} and \eqref{potential_Nf2}.
Finally, taking into account that matrix elements 
of the descendant operator $\widehat{L}_{-1}^{2}(z)V_{+}(z)$ 
between irregular vectors result in $\partial^{2}_{z}\Psi^{\bf i}_{\iota}$.
From eq.~\eqref{nullfield}, eqs.~\eqref{NullFMatrixEl_Nf12}
and eqs.~\eqref{L2} we get sought partial differential equations:
\begin{enumerate}
\addtocounter{equation}{1}
\item for ``single flavor'' degenerate irregular blocks:
\begin{equation}
\label{NVDeq1}
\left[\frac{1}{b^2}\,z^2\frac{\partial^2}{\partial z^2}-
\frac{4z}{3}\frac{\partial}{\partial z}+\frac{1}{4}z^2\Lambda^2 
+z\,m\,\Lambda+\frac{\Lambda^2}{z}
+\frac{\Lambda}{3}\frac{\partial}{\partial\Lambda}+
\frac{2\tilde\Delta+\Delta'-\Delta_+}{3}\right]\Psi^{\bf 1}_{\iota}\;=\;0\,,
\tag{\theequation D1}
\end{equation}

\item for ``two flavors'' degenerate irregular blocks:
\begin{multline}
\label{NVDeq2v2}
\left[\frac{1}{b^2}\,z^2\frac{\partial^2}{\partial z^2}-
\frac{3z}{2}\frac{\partial}{\partial z}
+\frac{1}{4}\Lambda^2\left(z^2+\frac{1}{z^2}\right)
+\Lambda\left(zm_{1}+\frac{m_{2}}{z}\right)\right.
\\
\left.+\frac{\Lambda}{2}\frac{\partial}{\partial\Lambda}+
\frac{\tilde\Delta+\Delta'-\Delta_+}{2}\right]\Psi^{\bf 2}_{\iota}\;=\;0\,.
\tag{\theequation D2}
\end{multline}

\end{enumerate}
In the next section we will consider the limit $b\to 0$ each of these equations
separately. The steps in forthcoming analysis follows directly those that 
has been already done in our previous work \cite{Piatek:2015jva}.
The experience gained in this study helps us 
to compute the semi-classical limit of eqs. \eqref{NVDeq1} and \eqref{NVDeq2v2}.

\section{Classical limit of \texorpdfstring{$N_f=1,2$}{Nf = 1,2} NVD equations}
\label{section3}

The differential equations we derived in the previous section are a 
starting point to the derivation of the differential equations some of which 
are well known in mathematics and physics. In what follows
we take the classical limit of NVD eqs.~\eqref{NVDeq1} and \eqref{NVDeq2v2}
as well as the functions that solve them. As steps leading to the equations
are the same in both cases $N_{f} = 1,2$ we discuss them in full generality.

The functions\footnote{Let us recall that indices I, II 
mean that different fusion rules, namely (\ref{fv2a}) and (\ref{fv2b}), 
have been assumed.}
$\Psi^{{\bf 1}}_{\iota}$, $\Psi^{{\bf 2}}_{\iota}$, $\iota={\rm I}, {\rm II}$
that solve eqs.~\eqref{NVDeq1}, \eqref{NVDeq2v2},
that are the degenerate irregular blocks given in eqs.~\eqref{P1} and \eqref{P2}
can be given explicit form by means of eqs.~\eqref{GaiottoNf=0}, \eqref{GaiottoNf=1}
and eq.~\eqref{CVO_normalization} which results in
\addtocounter{equation}{1}
\begin{multline}
\label{Phi1}
\Psi^{{\bf 1}}_{\iota}(z;\Lambda,m)
= z^{\Delta'-\Delta_+ -\tilde \Delta}
\sum_{r,s\geq 0}2^{-r}\Lambda^{2s+r}z^{r-s}\sum\limits_{p=0}^{[\frac{r}{2}]}\left(2m\right)^{r-2p}
\\
\times
\sum_{I\vdash r}\sum_{J\vdash s} \Big[G^{r}_{c,\Delta'}\Big]^{(2^p,1^{r-2p}) I}
\bra{\n_{\dl',I}}V_{+}(1)\ket{\n_{\tilde\dl,J}}
\Big[G^{s}_{c,\tilde\Delta}\Big]^{J (1^s)}
\tag{\theequation $\Psi1$}
\end{multline}
and
\begin{multline}
\label{Phi2}
\Psi^{{\bf 2}}_{\iota}(z;\Lambda,m_1, m_2) 
= z^{\Delta'-\Delta_+ -\tilde \Delta} \
\sum_{r,s\geq 0}(\tfrac{1}{2}\Lambda)^{r+s}z^{r-s}
\sum\limits_{p=0}^{[\frac{r}{2}]}\left(2m_1\right)^{r-2p}
\sum\limits_{p'=0}^{[\frac{s}{2}]}\left(2m_2\right)^{s-2p'}
\\
\times\sum_{I\vdash r}\sum_{J\vdash s} \Big[G^{r}_{c,\Delta'}\Big]^{(2^p,1^{r-2p}) I}
\bra{\n_{\dl',I}}V_{+}(1)\ket{\n_{\tilde\dl,J}}
\Big[G^{s}_{c,\tilde\Delta}\Big]^{J (2^{p'},1^{s-2p'})}.
\tag{\theequation $\Psi2$}
\end{multline}
In what follows it will be convenient to introduce the following notation
\begin{equation}
\label{Psi2Phi}
\Psi^{{\bf i}}_{\iota}(z;\Lambda,\,\cdot\,) \;=\; z^{\kappa_{\iota}}\,
\Phi^{\bf i}_{\iota}(z;\Lambda,\,\cdot\,)\;, 
\quad {\bf i}={\bf 1,2},
\quad \iota = {\rm I},\,{\rm II},
\end{equation} 
where 
\begin{eqnarray}\label{kappa}
\ka_{\iota} = \dl' - \dl_{+} - \tilde \dl = \left\{
    \begin{array}{rl}
      \Delta(\sigma+\tfrac{b}{4})-\Delta_{+}-\Delta(\sigma-\tfrac{b}{4})
      = -b\sigma-\Delta_{+} & \text{if } \iota=\text{I}
      \\[3pt]
      \Delta(\sigma-\tfrac{b}{4})-\Delta_{+}-\Delta(\sigma+\tfrac{b}{4}) 
      = b\sigma-\Delta_{+}\;\;\;  & \text{if } \iota=\text{II}.
    \end{array} \right.
\end{eqnarray}
Let us notice that  
$\Phi^{\bf i}_{\iota}(z;\Lambda,\,\cdot\,)$
can be split into ``diagonal'' $r=s$ and ``off-diagonal'' $r\neq s$ parts
$\Phi^{\bf i}_{\iota}(z;\Lambda,\,\cdot\,)=\Phi_{\iota, r=s}^{\bf i}(\Lambda,\,\cdot\,)
+\Phi_{\iota, r\neq s}^{\bf i}(z;\Lambda,\,\cdot\,)$. The former does not depend on $z$
leaving this dependence entirely to the latter. Making use of this observation
$\Psi^{\bf i}_{\iota}(z;\Lambda,\,\cdot\,)$ can be cast into the factorized form
\begin{equation}
\label{Psi12}
\Psi^{\bf i }_{\iota}(z;\Lambda,\,\cdot\,)\;=\;
z^{\kappa_{\iota}}
\,{\rm e}^{{\cal Y}^{\bf i}_{\iota}(\Lambda,\,\cdot\,)}
\,{\rm e}^{{\cal X}^{\bf i}_{\iota}(z;\Lambda,\,\cdot\,)},
\quad {\bf i}={\bf 1, 2},
\quad \iota = {\rm I},\,{\rm II}.
\end{equation}
The functions in the exponent take the form
\begin{equation*}
{\cal Y}^{\bf i}_{\iota}(\Lambda,\,\cdot\,)\;=\;\log\Phi_{\iota, r=s}^{\bf i}(\Lambda,\,\cdot\,),
\qquad
{\cal X}^{\bf i}_{\iota}(z;\Lambda,\,\cdot\,) \;
=\;\log\left(1+\frac{\Phi_{\iota, r\neq s}^{\bf i}(z;\Lambda,\,\cdot\,)}
{\Phi_{\iota, r=s}^{\bf i}(\Lambda,\,\cdot\,)}\right).
\end{equation*}
The factorized form in eq.~\eqref{Psi12}, as we soon see, gives one some insight into the behavior 
of $\Psi^{\bf i }_{\iota}$ in the classical limit. The equations \eqref{NVDeq1} and \eqref{NVDeq2v2}
under substitution \eqref{Psi2Phi} take the form
\begin{equation}
\label{before_limit}
\left[\frac{1}{b^2}\,z^2\frac{\partial^2}{\partial z^2}
+\left(\frac{2\kappa_{\iota}}{b^2}-\frac{5-\mathbf{i} }{4-\mathbf{i}}\right)z\frac{\partial}{\partial z}
+ \frac{\Lambda}{4-\mathbf{i}}\frac{\partial}{\partial\Lambda}
+ \mathcal{U}^{\bf i}_{\iota}(z;\Lambda,\,\cdot\,) \right] \Phi^{\bf i}_{\iota}(z;\Lambda,\,\cdot\,)=0\;,
\end{equation} 
where ${\bf i}={\bf 1, 2}$, $\iota = {\rm I},\,{\rm II}$ and
\begin{align*}
\mathcal{U}^{\bf 1}_{\iota}(z;\La,m) &= \frac{\kappa_{\iota}(\kappa_{\iota}-1)}{b^2}
-\frac{4\kappa_{\iota}}{3}+\frac{1}{4}z^2\Lambda^2 +
z\,m\,\Lambda + \frac{\Lambda^2}{z}
+\frac{2\tilde\Delta+\Delta'-\Delta_+}{3}\;,
\\[5pt]
\mathcal{U}^{\bf 2}_{\iota}(z;\La,m_{1},m_{2})&=\frac{\kappa_{\iota}(\kappa_{\iota}-1)}{b^2}
-\frac{3\kappa_{\iota}}{2}+\,\frac{1}{4}\Lambda^2 \left(z^2+\frac{1}{z^2}\right)
+\Lambda\left(zm_1+\frac{m_2}{z}\right)
+ \frac{\tilde\Delta+\Delta'-\Delta_+}{2}.
\end{align*}

Let us consider the limit $b\to 0$ of eqs.~\eqref{before_limit}.
To this purpose it is convenient to replace the parameter $\sigma$ in 
$\dl'$ and $\tilde\dl$ with $\xi=b\sigma$, cf.~(\ref{Delta_sigma}).
Recall that the weights $\dl'$, $\tilde\dl$ are related by the fusion rules
written in eqs.~(\ref{fv2a}) and (\ref{fv2b}). 
Hence, in the limit $b\to 0$
\begin{itemize}
\item[---] the conformal weights in eqs.~\eqref{before_limit} read as follows
\begin{eqnarray*}
\label{delta}
\Delta',\tilde\Delta &\stackrel{b\to 0}{\sim}& \frac{1}{b^2}\,\delta,
\qquad
\text{where}
\qquad
\delta=\lim_{b\to 0}b^2\Delta'=\lim_{b\to 0}b^2\tilde\Delta=\tfrac{1}{4}-\xi^2,
\\
\Delta_{+} &\stackrel{b\to 0}{\sim}& {\cal O}(b^0)
\qquad
\text{and}
\qquad
2\tilde\Delta+\Delta'-\Delta_+\stackrel{b\to 0}{\sim}\frac{1}{b^2}\,3\delta,
\qquad
\tilde\Delta+\Delta'-\Delta_+\stackrel{b\to 0}{\sim}\frac{1}{b^2}\,2\delta;
\end{eqnarray*}
\item[---] the $\kappa_{\iota}$'s yield (cf.~(\ref{kappa}))
\begin{eqnarray*}
\ka_{\iota} &\stackrel{b\to 0}{\longrightarrow}& \left\{
    \begin{array}{rl}
     -\xi+\tfrac{1}{2} & \text{if } \iota=\text{I}
      \\[3pt]
      \xi+\tfrac{1}{2} & \text{if } \iota=\text{II}
    \end{array} \right.
\\  
\kappa_{\iota}\left(\kappa_{\iota}-1\right) &\stackrel{b\to 0}{\longrightarrow}&
-\left(\tfrac{1}{4}-\xi^2\right)\;=\;-\delta
\qquad 
\text{for}\;\iota=\text{I, II}.
\end{eqnarray*}
\end{itemize}
Making use of the relationship between CFT and instanton parameters
given in eqs.~\eqref{para1}, \eqref{para2} that reveals their $b$ dependence
i.e., $\Lambda=\hat\Lambda/(\epsilon_1 b)$, $m = m_1=\hat m_1/(\epsilon_1 b)$,
$m_2=\hat m_2/(\epsilon_1 b)$, and taking into account the fact 
that the external conformal weights $\Delta'$, $\tilde\Delta$ are heavy 
in the classical limit, while the degenerate weight $\dl_{+}$ labeling the degenerate vertex operator
is light, we conjecture that the degenerate irregular blocks given in eqs.~\eqref{P1} and \eqref{P2} 
factorize into the light and the heavy parts in the classical limit. 
For $\Phi^{\bf i}_{\iota}=z^{-\kappa_{\iota}}\Psi^{\bf i}_{\iota}$
this assertion translates into the following asymptotic behavior in the limit $b\to 0$
\begin{equation}
\label{factorization}
\Phi^{\bf i}_{\iota}(z;\hat\Lambda/(\epsilon_1 b),\,\cdot\,)
\stackrel{b\to 0}{\sim}
\varphi^{\bf i}_{\iota}\!\left( z;\hat\Lambda/\epsilon_1,\;\cdot\;\right)\,
{\rm e}^{\frac{1}{b^2}f_{\delta}^{\bf i}\left(\hat\Lambda/\epsilon_1,\;\cdot\;\right)} ,
\qquad
{\bf i}={\bf 1,2},
\qquad 
\iota=\text{I, II}.
\end{equation} 
Comparing the right hand sides of eqs.~\eqref{factorization} to eqs.~\eqref{Psi12} 
one can expect that 
\addtocounter{equation}{1}
\begin{align}
\label{light_factor}
\varphi^{\bf i}_{\iota}\!\left( z;\hat\Lambda/\epsilon_1,\;\cdot\;\right) &=
\lim\limits_{b\to 0}{\rm e}^{{\cal X}^{\bf i}_{\iota}( z;\Lambda,\;\cdot\;)}
=\lim\limits_{b\to 0}
\left(1+\frac{\Phi_{\iota, r\neq s}^{\bf i}( z;\hat\Lambda/(\epsilon_1 b),\;\cdot\;)}
{\Phi_{\iota, r=s}^{\bf i}(\hat\Lambda/(\epsilon_1 b),\;\cdot\;)}\right),
\tag{\theequation $\varphi$}
\\[5pt]
\label{heavy_factor}
f_{\delta}^{\bf i}\!\left(\hat\Lambda/\epsilon_1,\;\cdot\;\right) &= 
\lim\limits_{b\to 0}b^2 {\cal Y}^{\bf i}_{\iota}(\Lambda,\;\cdot\;)
= \lim\limits_{b\to 0}b^2\log\Phi_{\iota, r=s}^{\bf i}\left(\hat\Lambda/(\epsilon_1 b),\;\cdot\;\right)
\qquad 
\text{for}\;\iota=\text{I, II}.
\tag{\theequation $f$}
\end{align}
In ref. \cite{Piatek:2015jva}, where the similar discussion was 
performed for the $N_{f}=0$ case to obtain the Mathieu equation, we were able to prove this
conjecture keeping only dominating factors in $b^{-2}$ and neglecting all the sub-dominating ones.
Although much complex, a similar proof, in principle, could be performed in the case under study.
We expect, however, similar results, therefore, in what follows, we are content with 
the assumption of the factorization \eqref{factorization} as a fact.

Taking into account the factorization \eqref{factorization}, in the limit $b\to 0$
from eqs.~\eqref{before_limit} one gets 
\begin{itemize}
\item[---] for $\iota={\rm I}$:
\begin{equation}
\label{post_limit_I}
\sbr{
z^2\frac{\textrm{d}^2}{\textrm{d} z^2}+2(\tfrac{1}{2}-\xi)z\frac{\textrm{d}}{\textrm{d} z}
+ \mathscr{U}^{\mathbf{i} }(z;\hat\Lambda/\epsilon_1,\;\cdot\;)
+ \frac{\hat\Lambda}{4-\mathbf{i}}\,\partial_{\hat\Lambda}f_{\delta}^{\bf i}(\hat\Lambda/\epsilon_1,\;\cdot\;)
}\vf^{\mathbf{i}}_{\rm I}(z;\hat\Lambda/\epsilon_1,\;\cdot\;) = 0,
\end{equation}
\item[---] for $\iota={\rm II}$:
\begin{equation}
\label{post_limit_II}
\sbr{
z^2\frac{\textrm{d}^2}{\textrm{d} z^2}+2(\tfrac{1}{2}+\xi)z\frac{\textrm{d}}{\textrm{d} z}
+ \mathscr{U}^{\mathbf{i} }(z;\hat\Lambda/\epsilon_1,\;\cdot\;)
+ \frac{\hat\Lambda}{4-\mathbf{i}}\,\partial_{\hat\Lambda}f_{\delta}^{\bf i}(\hat\Lambda/\epsilon_1,\;\cdot\;)
}\vf^{\mathbf{i}}_{\rm II}(z;\hat\Lambda/\epsilon_1,\;\cdot\;) = 0,
\end{equation}
\end{itemize}
where $\mathbf{i} = \mathbf{1,2}$ and
\begin{align}
\label{U1scri}
\mathscr{U}^{\mathbf{1} }\!\cbr{z;\tfrac{\hat\Lambda}{\epsilon_1},m}
&=\frac{1}{4}\frac{\hat\Lambda^2}{\epsilon_1^2}\,z^2 
+\frac{\hat\Lambda\hat m}{\epsilon_1^{2}}\,z
+\frac{\hat\Lambda^2}{\epsilon_1^2}\,\frac{1}{z}\;,
\tag{\theequation $\mathscr{U}1$}
\\
\label{U2scri}
\mathscr{U}^{\mathbf{2} }\!\cbr{z;\tfrac{\hat\Lambda}{\epsilon_1},m_{1},m_{2}}
&=\frac{1}{4}\frac{\hat\Lambda^2}{\epsilon_1^2}\left(z^2+\frac{1}{z^2}\right)
+\frac{\hat\Lambda}{\epsilon_1}
\left(\frac{\hat m_1}{\epsilon_1}\,z +\frac{\hat m_2}{\epsilon_1}\,\frac{1}{z}\right)\;.
\tag{\theequation $\mathscr{U}2$}
\end{align}
The classical blocks $f_{\delta}^{\bf 1},\,f_{\delta}^{\bf 2}$ are defined 
in eqs. \eqref{ClIrrblock1} and \eqref{ClIrrblock2}. Deriving the above equations \eqref{post_limit_I}
and \eqref{post_limit_II}
we assumed that $b^{2}\hat\Lambda\partial_{\hat\Lambda}\varphi^{\bf i}_{\iota}\to 0$ for $b\to 0 $.
This conjecture was verified by direct computations of $\hat\Lambda/\epsilon_1$ expansion of 
$\varphi^{\bf i}_{\iota}$ 
(see below). Although it has not been proved, in analogy to the $N_{f}=0$ case in ref. \cite{Piatek:2015jva}, 
where it was proved up to leading order in $b^{-2}$ that $\vf^{\mathbf{0}}$ is independent of $b$ ,
it is here expected that $\vf^{\mathbf{i}}_{\iota}$ also does not depend on $b$ 
as well as its $\hat\La$ derivative. 

In order to get rid of the first order 
differential operator in eqs.~\eqref{post_limit_I}
and \eqref{post_limit_II} we redefine the functions $\vf^{\mathbf{i}}_{\rm I}$, $\vf^{\mathbf{i}}_{\rm II}$:
\begin{equation}
\label{solutions}
\vf^{\mathbf{i}}_{\rm I} \;=\; z^{\xi}\psi^{\mathbf{i}}_{\rm I},
\;\;\;\;\;\;\;\;\;\;\;\;\;\;
\vf^{\mathbf{i}}_{\rm II} \;=\; z^{-\xi}\psi^{\mathbf{i}}_{\rm II}
\end{equation}
and subsequently change a variable $z=\ex^{w}$. In result we obtain 
\begin{equation}
\label{eq_class_lim_final}
\sbr{
\frac{\textrm{d}^{2}}{\textrm{d} w^{2}} + \mathscr{U}^{\mathbf{i}}(\ex^{w};\hat\La/\e_{1},\;\cdot\;) 
+ \frac{\hat\Lambda}{4-\mathbf{i}}\,\partial_{\hat\Lambda}f_{\delta}^{\bf i}(\hat\Lambda/\epsilon_1,\;\cdot\;) - \xi^{2} 
}
\psi^{\mathbf{i}}_{\rm I}(\ex^{w};\hat\La/\e_{1},\;\cdot\;) = 0\,,
\quad \mathbf{i} = \mathbf{1,2}\;
\end{equation} 
and the same equations for the second solutions $\psi^{\mathbf{i}}_{\rm II}(\ex^{w};\hat\La/\e_{1},\;\cdot\;)$.
Potentials $\mathscr{U}^{\mathbf{i}}$ are defined in eqs. \eqref{U1scri}, \eqref{U2scri}.
Thus we have obtained two complex second order Schr\"{o}dinger type differential equations with spectra 
$\la^{\mathbf{i}}_{\xi}/4 :=\xi^{2} - \hat\La\partial_{\hat\Lambda}f_{\delta}^{\bf i}/(4-\mathbf{i})$,
$\mathbf{i} = \mathbf{1,2}$ and pairs $(\psi^{\mathbf{i}}_{\rm I}, \psi^{\mathbf{i}}_{\rm II})$ 
of independent solutions.
Below we address each of the case $N_{f} = 1$ and $N_{f}=2$ separately,
where we compute the spectra and eigenfunctions of relevant operators.

\subsection{Single flavor case: solvable complex potential}
In this subsection we narrow down the discussion to $N_{f}=1$ case. 
In order to obtain the differential equation for this case 
that takes the form of Hill's equation \eqref{HillEq}, but that possesses 
a complex periodic potential,
we identify the ``coupling constant'' as $h=2\hat m/\epsilon_1=2\hat\Lambda/\epsilon_1$ 
in eq.~\eqref{eq_class_lim_final} with potential~\eqref{U1scri} and, 
after the change of variable $w = -2 i x$, we obtain\footnote{A reason
we choose this particular change of variable is dictated by experience gained in ref. \cite{Piatek:2015jva},
where this choice led to the coincidence of the expansion of pure gauge three-point degenerate irregular block 
with similar ``weak coupling''  expansion of the Mathieu exponent $\mathrm{me}_{\n}(x)$.}
\begin{equation}
\label{HillsEqNf1}
\left[-\frac{{\rm d}^2}{{\rm d}x^2}+\frac{h^2}{4}\,{\rm e}^{-4 i x}+2 h^2\cos2x\right]
\psi^{\bf 1}
=\lambda_{\xi}^{\mathbf{1}}\,\psi^{\bf 1}\;,
\qquad
x\in \mathbb{R}/\pi\mathbb{Z}.
\end{equation}
This is the Schr\"{o}dinger equation with complex odd potential on unit circle.
The eigenvalue $\lambda_{\xi}^{\mathbf{1}}$ can be obtained from eqs. \eqref{ClIrrexp2} and \eqref{f1coeffs}
with parametric relation to Floquet's exponent $\xi=\n/2$,
\begin{equation}
\label{spectrumNf1}
\begin{aligned}
\lambda_{\n}^{\mathbf{1}}(h) &= \n^2-\frac{4 h}{3}\pt_{h} 
f_{\frac{1}{4}(1-\n^2)}^{\bf 1}\!\left(\tfrac{1}{2}h,\tfrac{1}{2}h\right)
\\
&=\n^2
+\frac{2}{3(\nu^{2} -1) }h^{4} 
+\frac{3}{32 (\nu^{2} -4) (\nu^{2} -1)}h^{6}
+\frac{5 \nu ^2+7}{8 (\nu^{2} -4) (\nu^{2} -1)^3}h^{8}+\ord{h^{10}}.
\end{aligned}
\end{equation}
The expression (\ref{spectrumNf1}) is well defined for $\nu\notin\mathbb{Z}$.
The first few terms in the expansion (\ref{spectrumNf1}) suggest that the spectrum $\lambda_{\n}^{\mathbf{1}}(h)$
is real for $h\in\mathbb{R}$ and $\nu\in\mathbb{R}\setminus\mathbb{Z}$. 
Indeed, from our approach easily follows that the quantity $\lambda_{\n}^{\mathbf{1}}(h)$ is 
real-valued for $b, \hat\Lambda, \hat m, \epsilon_1, \xi\in\mathbb{R}$. 
The eigenvalue $\lambda_{\n}^{\mathbf{1}}(h)$ in 
eq.~(\ref{spectrumNf1}) is given by the logarithmic derivative of the $N_f=1$ classical irregular block:
$$
f_{\delta}^{\bf 1}\!\left(\hat\Lambda/\epsilon_1, \hat m/\epsilon_1\right)=
\lim\limits_{b\to 0} b^2 \log {\cal F}_{1+6Q^2, \frac{1}{b^2}\delta}
\!\left(\hat\Lambda/(\epsilon_1 b), \hat m/(\epsilon_1 b)\right).
$$
Let us recall that the $N_f=1$ quantum irregular block ${\cal F}_{c, \Delta}\!\left(\Lambda, m\right)$
is defined as an expansion in $\Lambda$ and $m$ with coefficients which are elements
of the inverse of the Gram matrix, i.e.~rational functions of $c$ and $\Delta$.
Therefore, ${\cal F}_{c, \Delta}\!\left(\Lambda, m\right)$
is real-valued function when all its parameters: 
$$
c=1+6Q^2,
\;\;\;\;\;\;\;\;
\Delta=\frac{1}{b^2}\delta=\frac{1}{b^2}(\tfrac{1}{4}-\xi^2), 
\;\;\;\;\;\;\;\;
\Lambda=\hat\Lambda/(\epsilon_1 b), 
\;\;\;\;\;\;\;\;
m=\hat m/(\epsilon_1 b)
$$
are real. Hence, in particular, for  $b, \hat\Lambda, \hat m, \epsilon_1, \xi\in\mathbb{R}$
and ${\cal F}_{c, \Delta}\!\left(\Lambda, m\right)\geq0$ the function
$\log{\cal F}_{c, \Delta}\!\left(\Lambda, m\right)$ is real-valued.
If ${\cal F}_{c, \Delta}\!\left(\Lambda, m\right)<0$ then the function 
$\log{\cal F}_{c, \Delta}\!\left(\Lambda, m\right)$ will have constant imaginary part $i\pi$
which vanishes in the limit $b\to0$.

In our approach the corresponding eigenfunctions, i.e.~two 
independent solutions of eq.~\eqref{HillsEqNf1} can be 
given in the ``weak coupling''  expansion (small $h$). The eigenfunctions are 
calculable from the classical limit of the ``off-diagonal'' part
of $\Phi_{\iota, r\neq s}^{\bf 1}$ given in eq.~\eqref{Phi1} (see also eqs.~\eqref{light_factor} and 
(\ref{solutions})) 
in the same manner as in ref. \cite{Piatek:2015jva}. As a result one finds\footnote{Here, $\xi=\n/2$.}
\begin{subequations}
\begin{eqnarray}
\label{eigenfuncNf1}
\psi^{\bf 1}_{\rm I}\;\equiv\;\psi^{\bf 1}_{\rm I}\!
\left({\rm e}^{-2 i x};\tfrac{1}{2}h,\tfrac{1}{2}h\right) 
&=& {\rm e}^{i \n x}\,
\lim\limits_{b\to 0}
\left(1+\frac{\Phi_{{\rm I}, r\neq s}^{\bf 1}({\rm e}^{-2 i x};\frac{h}{2b},\frac{h}{2b})}
{\Phi_{{\rm I}, r=s}^{\bf 1}(\frac{h}{2b},\frac{h}{2b})}\right)\nonumber
\\
&=& {\rm e}^{i \n x} + \sum_{n\geq 1} \mathcal{R}^{\mathbf{1}}_{{\rm I}, n}(x;\n)\cbr{\frac{h}{2}}^{2 n}\;,
\end{eqnarray}
where three first coefficients are found to take the form
\begin{equation}
\label{eigenfuncNf1_coeffs}
\begin{aligned}
\mathcal{R}^{\mathbf{1}}_{{\rm I}, 1}(x;\n) & =
\frac{\ex^{(\nu -2) ix}}{\nu -1}
-\frac{\ex^{(\nu +2) ix}}{\nu +1}
+\frac{\ex^{ (\nu -4)i x}}{8 (\nu -2)}\; , 
\\
\mathcal{R}^{\mathbf{1}}_{{\rm I}, 2}(x;\n) & = \frac{1}{2}\left(
\frac{(5 - \nu ) \ex^{ (\nu -2)i x}}{4 (\nu -2) (\nu^{2}-1)}
+\frac{\ex^{(\nu -4) ix}}{(\nu -2) (\nu -1)}
+\frac{\ex^{(\nu +4) ix}}{(\nu +1)(\nu +2)}
-\frac{(5-3\nu ) \ex^{ (\nu -6)i x}}{12(\nu -3) (\nu -2) (\nu -1)}
\right.
\\
& \qquad\left.
+\frac{\ex^{(\nu -8)i x }}{64 (\nu -4) (\nu - 2)}
\right),
\\
\mathcal{R}^{\mathbf{1}}_{{\rm I}, 3}(x;\n) & = \frac{1}{2}\left(
\frac{\left(\nu ^2-4 \nu +7\right) \ex^{(\nu -2) ix}}{(\nu -2) (\nu -1)^3 (\nu+1)}
-\frac{\left(\nu ^2+4 \nu +7\right) \ex^{(\nu +2) ix}}{(\nu -1) (\nu +1)^3 (\nu +2)}
+ \frac{2 \ex^{ (\nu -4)i x}}{3 (\nu -3) (\nu -2) (\nu^{2} -1)}
\right.
\\
&\qquad
\left.
-\frac{\ex^{(\nu +6) ix}}{3(\nu +1) (\nu +2) (\nu +3)}
+\frac{\ex^{(\nu -6) ix}}{3 (\nu -3) (\nu -2) (\nu -1)}
\right.
\\
&
\qquad
\left. 
+ \frac{(9-\nu ) \ex^{ (\nu -6)i x}}{64 (\nu -4) (\nu -2) (\nu -1) (\nu +1)}
+\frac{(3 \nu -7) \ex^{(\nu -8)i x}}{6 (\nu -4) (\nu -3) (\nu -2) (\nu -1)}
\right.
\\
&
\qquad
\left. 
+\frac{\left(15 \nu ^2-80 \nu +89\right) \ex^{ (\nu -10)i x}}{960 (\nu -5) (\nu -4) (\nu -3) (\nu
   -2) (\nu -1)}+\frac{\ex^{(\nu -12)i x }}{1536 (\nu -6) (\nu -4) (\nu -2)}
\right)\;,
\end{aligned}
\end{equation}
\end{subequations}
and so on. Analogously, the second solution that solves eq.~\eqref{HillsEqNf1} 
is given by the formula:
\begin{equation}
\label{eigenfuncNf1_II}
\psi^{\bf 1}_{\rm II}\;\equiv\;\psi^{\bf 1}_{\rm II}\!
\left({\rm e}^{-2 i x};\tfrac{1}{2}h,\tfrac{1}{2}h\right) 
= {\rm e}^{-i \n x}\,
\lim\limits_{b\to 0}
\left(1+\frac{\Phi_{{\rm II}, r\neq s}^{\bf 1}({\rm e}^{-2 i x};\frac{h}{2b},\frac{h}{2b})}
{\Phi_{{\rm II}, r=s}^{\bf 1}(\frac{h}{2b},\frac{h}{2b})}\right).
\end{equation}
We have found no examples in the literature
for this equation and its solutions which we could compare with. 

\subsection{Two flavors case: Whittaker--Hill equation}
In $N_{f}=2$ case, after change of variable $w=-2ix$ in eq.~\eqref{eq_class_lim_final} 
for $\mathbf{i}=2$ with potential \eqref{U2scri}, 
and identifying the ``coupling constant'' $h=2\hat\Lambda/\epsilon_1$, and
introducing the second parameter $\m = \hat m_1/\epsilon_1 = \hat m_2/\epsilon_1$
the equation assumes the form:
\begin{equation}
\label{Whittaker_Hill_Eq}
\left[-\frac{{\rm d}^2}{{\rm d}x^2}+\frac{1}{2}h^2\cos 4x + 4h\mu\cos 2x\right]
\psi^{\bf 2}
=\lambda^{\mathbf{2}}_{\xi} \,\psi^{\bf 2}\;,
\end{equation}
which is a specific example of the Hill equation \eqref{HillEq} termed the Whittaker--Hill equation.
The eigenvalue $\lambda^{\bf 2}_{\xi}$, after the identification $\xi=\n/2$, 
is given by the following formula (cf.~\eqref{ClIrrexp3} and \eqref{f2coeffs})
\begin{multline*}
\lambda^{\mathbf{2}}_{\n}(h,\m) 
= \n^2-2 h\,\pt_{h}f_{\frac{1}{4}(1-\n^2)}^{\bf 2}\!\left(\tfrac{1}{2}h,\m,\m\right)
\\
=\n^{2} + \frac{2}{\nu ^2-1}\mu ^2h^2 
+\cbr{\frac{1}{2(\nu ^2-4)}
+\frac{12 }{(\n^{2}-1)(\n^{2}-4)}\mu ^2
   +\frac{8(5 \nu ^2+7)}{\left(\nu ^2-4\right) \left(\nu^2-1\right)^3}\mu ^4
}\frac{h^{4}}{16}
\\
+\left(
\frac{20 }{\left(\nu^2-9\right) \left(\nu^2-4\right) \left(\nu^2-1\right)}\mu^2
+\frac{32\left(7 \nu ^2+17\right)}{\left(\nu ^2-9\right) \left(\nu
   ^2-4\right) \left(\nu ^2-1\right)^3}\mu ^4
\right. 
\\
\left. 
+\frac{64 \left(9 \nu ^4+58 \nu ^2+29\right)}{\left(\nu ^2-9\right)
   \left(\nu ^2-4\right) \left(\nu ^2-1\right)^5}\mu ^6 
\right)\frac{h^{6}}{64}
+\ord{(h/2)^{8}}
\;.
\end{multline*}
Hence, as before the 2dCFT technics led to the so-called non-integer order 
($\nu\notin\mathbb{Z}$) solution of the Whittaker--Hill equation. 
The above expression for $\lambda^{\mathbf{2}}_{\n}$ 
exactly coincides with that obtained in ref.~\cite{Rim:2015tsa} by means of the perturbation 
calculus (see \cite{MuellerKirsten:2006}) applied to eq.~\eqref{Whittaker_Hill_Eq}.

As in the previous case, the two independent solutions of eq.~\eqref{Whittaker_Hill_Eq} can be 
given in the ``weak coupling'' expansions and are computed from the classical limit of the ``off-diagonal'' part
of $\Phi_{\iota, r\neq s}^{\bf 2}$ given in eq.~\eqref{Phi2} (see also eqs.~\eqref{light_factor} and 
(\ref{solutions})).
As a result one finds
\begin{subequations}
\begin{multline}
\label{eigenfuncNf2}
\psi^{\bf 2}_{\rm I}\;\equiv\;\psi^{\bf 2}_{\rm I}\!\left({\rm e}^{-2ix};\tfrac{1}{2}h,\mu,\mu \right) 
\\
= {\rm e}^{ix\n}\;
\lim\limits_{b\to 0}
\left(1+\frac{\Phi_{{\rm I}, r\neq s}^{\bf 2}({\rm e}^{-2ix};\frac{h}{2b},\frac{\mu}{b},\frac{\mu}{b})}
{\Phi_{{\rm I}, r=s}^{\bf 2}\left(\frac{h}{2b},\frac{\mu}{b},\frac{\mu}{b}\right)}\right)
= \ex^{i\n x} + \sum_{n\geq 1}\mathcal{R}^{\mathbf{2}}_{{\rm I},n}(x;\n,\m)\cbr{\frac{h}{2}}^{2n},
\end{multline}
where the first tree coefficients of the Whittaker-Hill function take the form
\begin{equation}
\label{eigenfuncNf2_coeffs}
\begin{aligned}
\mathcal{R}^{\mathbf{2}}_{{\rm I}, 1}(x;\n,\m) &= \m\cbr{\frac{{\rm e}^{-i x (\nu +2)}}{\nu -1}
-\frac{{\rm e}^{-i x (\nu -2)}}{\nu +1}}\;,
\\
\mathcal{R}^{\mathbf{2}}_{{\rm I}, 2}(x;\n,\m) &= 
\frac{{\rm e}^{-i x (\nu +4)}}{8 (\nu -2)}
-\frac{{\rm e}^{-i x (\nu -4)}}{8 (\nu +2)}
+\frac{1}{2} \mu ^2 \left(\frac{{\rm e}^{-i x (\nu -4)}}{(\nu +1) (\nu +2)}+\frac{{\rm e}^{-i x (\nu
   +4)}}{(\nu -2) (\nu -1)}\right) \;,
\\
\mathcal{R}^{\mathbf{2}}_{{\rm I}, 3}(x;\n,\m)&= \m\left(
-\frac{(\nu -5) {\rm e}^{-i x (\nu +2)}}{8 (\nu -2) (\nu -1) (\nu +1)}-\frac{(\nu +5) {\rm e}^{-i x (\nu
   -2)}}{8 (\nu -1) (\nu +1) (\nu +2)}
   \right. 
\\
&\hskip 3cm\left. +\frac{(3 \nu -5) {\rm e}^{-i x (\nu +6)}}{24 (\nu -3) (\nu -2)
   (\nu -1)}+\frac{(3 \nu +5) {\rm e}^{-i x (\nu -6)}}{24 (\nu +1) (\nu +2) (\nu +3)}
\right)
\\
&\quad+ \m^{3}\left(
\frac{\left(\nu ^2-4 \nu +7\right) {\rm e}^{-i x (\nu +2)}}{2 (\nu -2) (\nu -1)^3 (\nu
   +1)}
   -\frac{\left(\nu ^2+4 \nu +7\right) {\rm e}^{-i x (\nu -2)}}{2 (\nu -1) (\nu +1)^3 (\nu
   +2)}
   \right. 
\\
&\hskip 3cm\left. 
   +\frac{{\rm e}^{-i x (\nu +6)}}{6 (\nu -3) (\nu -2) (\nu -1)}
   -\frac{{\rm e}^{-i x (\nu -6)}}{6 (\nu
   +1) (\nu +2) (\nu +3)}
\right)\;.
\end{aligned}
\end{equation} 
\end{subequations}
In order to obtain the second solution it suffices to take  
\begin{equation}
\label{eigenfuncNf2_II}
\psi^{\bf 2}_{\rm II}\;\equiv\;\psi^{\bf 2}_{\rm II}
\left({\rm e}^{-2ix};\frac{h}{2b},\frac{\mu}{b},\frac{\mu}{b}\right) 
= {\rm e}^{-i \n x}\,
\lim\limits_{b\to 0}
\left(1+\frac{\Phi_{{\rm II}, r\neq s}^{\bf 2}({\rm e}^{-2ix};\frac{h}{2b},\frac{\mu}{b},\frac{\mu}{b})}
{\Phi_{{\rm II}, r=s}^{\bf 2}(\frac{h}{2b},\frac{\mu}{b},\frac{\mu}{b})}\right).
\end{equation}
In analogy to 
the Mathieu's exponent one can term the above solutions generalized Mathieu exponents
or Whittaker-Hill exponents. 

\section{Conclusions}
\label{section4}
In the present paper we have shown that the $N_f=1$ and $N_f=2$ 
classical irregular blocks solve the Schr\"{o}dinger eigenvalue problem for
the complex (\ref{complexQ}) and Whittaker--Hill (\ref{WH}) potentials, respectively.\footnote{
Such observation in the case of the Whittaker--Hill operator has been
already made in \cite{Rim:2015tsa}. 
The $N_f=1$ case has not been discussed in the literature so far.}
In addition, for each of the above cases we have derived within CFT 
framework corresponding two linearly independent eigenfunctions. 
These eigenfunctions are determined by two different 
fusion rules imposed on the $in$- and $out$-states conformal weights
appearing in the degenerate irregular blocks, in accordance with the Feigin--Fuchs Theorem.

The conformal field theory setup proved to be especially powerful in the 
study of the eigenvalue problem for the complex potential (\ref{complexQ}).
For $x\in\mathbb{R}$ the potential (\ref{complexQ})
is $\pi$-periodic and the corresponding quantum-mechanical hamiltonian is
evidently {\sf PT}-symmetric. It turned out that such hamiltonian
in the weak coupling region (small $h\in\mathbb{R}$) and for the real
non-integer values of the Floquet parameter $\nu\notin\mathbb{Z}$
has a real spectrum determined by the $N_f=1$ classical irregular block expansion. 
From the definition of the $N_f=1$ classical irregular block immediately
follows that the spectrum is real for $h\in\mathbb{R}$ and 
$\nu\in\mathbb{R}\setminus\mathbb{Z}$. Therefore, we have found
yet another new example of the {\sf PT}-invariant hamiltonian with periodic potential
which has a real spectrum. Such hamiltonians have interesting applications in the branch of 
the condensed matter theory known as ``complex crystals'' (see e.g.~\cite{Bender:1998uc,Bender:2007nj}) 
or more in general provide models for testing
postulates of the {\sf PT}-symmetric quantum mechanics (cf.~\cite{Bender:2007nj} and refs.~therein).

As has been already mentioned our solutions to eqs.~(\ref{HillsEqNf1}) and 
(\ref{Whittaker_Hill_Eq}) make sense for small $h\in\mathbb{R}$ 
and $\nu\notin\mathbb{Z}$. Hence, two interesting questions arise at 
this point:~(i) How within 2dCFT one can get the solutions in the other regions of the
spectrum, in particular, for large coupling constant(s)? (ii) How is it
possible to derive from the irregular blocks the solutions with integer values of the Floquet
parameter? Work is in progress in order to answer these questions.

As final remark let us stress that it is possible to extend
results of the present paper to generic situation when $N_f=$ odd/even number of flavors,
cf.~\cite{Bonelli:2011aa, Rim:2015tsa}. 
The case with an even number of flavors should lead to the solution of the 
equation termed in \cite{Rim:2015tsa} as ``generalized Mathieu equation'', i.e.~Schr\"{o}dinger equation
with potential built out of higher cosine terms. In the same way the NVD equations obeyed by the 
three-point degenerate irregular blocks with an odd number of flavors will produce in the
classical limit solutions of the Schr\"{o}dinger equation with potential being generalization of
(\ref{complexQ}). The latter case seems to be an interesting task for further investigation due its 
possible application as a laboratory to study implications of {\sf PT} symmetry.

\appendix
\section{Floquet's theorem and band structure of spectrum}
\renewcommand{\theequation}{A.\arabic{equation}}
\setcounter{equation}{0}
\label{sec:App_A}

In this appendix we recall most important facts concerning 
the eq. \eqref{HillEq}. Let $y_{i}(x) = y_{i}(x;\la) ,\ i=1,2$ be the normalized fundamental
solutions of \eqref{HillEq}, i.e., such that $y_{1}(0) = y'_{2}(0)=1$ and
$y_{2}(0)=y'_{1}(0)=0$. The periodicity of $Q(x)$ implies that $y(x+\pi)$ is also
a solution of eq. \eqref{HillEq}. The fundamental solutions $y_{i}(x),\ i=1,2$
satisfy the following equation
\begin{equation}
\label{monodromy}
v(x+\pi) = v(x) M(\la),
\quad
v = (y_{1}\  y_{2}),
\qquad
M(\la) :=  
\begin{pmatrix}
y_{1}(\pi;\la) & y_{2}(\pi;\la)
\\
y_{1}'(\pi;\la) & y_{2}'(\pi;\la)
\end{pmatrix}
.
\end{equation} 
where $M(\la)$ is the monodromy matrix. Wronskian for the normalized fundamental solutions 
amounts to $W(y_{1},y_{2})=1$ which entails that $M(\la)\in SL(2,\mathbb{C})$. 
Since $y(x)$ is a linear combination of fundamental solutions 
and $y(x+\pi)$ also belongs to the set of all solutions to eq. \eqref{HillEq} it is possible to pick 
such one, that $y(x+\pi) = \r y(x)$. This amounts to the diagonalisation of monodromy matrix 
\begin{gather*}
\chi(\r):=\det\left(M(\la) - \r I\right)=\r^{2} - \dl(\la)\r + 1,
\\
\chi(\r_{1}) = \chi(\r_{2}) = 0,
\quad \r_{1} : = \ex^{i \n \pi},\ \r_{2} : = \ex^{-i \n \pi},
\quad \n \in \mathbb{C},
\end{gather*}
where $\dl(\la) := \mbox{tr}\,M(\la)$ is the \emph{Hill's discriminant},
$\r_{1},\,\r_{2}$ are the \emph{Floquet multipliers}
and $\n$ is the \emph{Floquet's exponent}. 
\begin{quotation}
\noindent
\textbf{Theorem}(Floquet) Let $\r_{1},\r_{2}$ be the roots of $\chi(\r)$.
\begin{enumerate}
\item If $\r_{1}\neq\r_{2}$ then 
$$
f_{1}(x) = \ex^{i\n x} p_{1}(x), \quad f_{2}(x) = \ex^{-i\n x} p_{2}(x), 
\qquad p_{i}(x+\pi) = p_{i}(x), \ i=1,2 
$$
are the two that span the space of solutions to Hill's equation \eqref{HillEq}.
\item  If $\r_{1} = \r_{2}$ the monodromy matrix $M(\la)$ is similar to upper-triangular matrix,
that is there exists a solution $p(x)$
which is either periodic or anti-periodic ($p(x+\pi)=-p(x)$) and the second linearly 
independent solution $q(x)$ such, that
$$
q(x+\pi) = \r_{1} q(x) + \a p(x).
$$
In case $\a=0$ the monodromy matrix $M(\la)\in\mathbb{Z}_{2}$, i.e., $y_{2}(\pi)=y'_{1}(\pi)=0$ and
all solutions are either periodic or anti-periodic.
\end{enumerate}
\end{quotation}
Note, that if $\n\in\mathbb{C}$ with $\Im{\n}\neq 0$ the solutions are unbounded to which one refers as 
to \emph{unstable} solutions. In order to have \emph{stable} solutions it is necessary for the 
Floquet exponent $\n\in\mathbb{R}$. However, it is insufficient condition as follows from the first part of the second
point of the Floquet theorem. A sufficient condition is $\n\in\mathbb{R}$ and $M(\la)\in\mathbb{Z}_{2}$ which 
entails that $\n =1,2$. In this case all solutions are stable and periodic with the basic period $\pi$ or $2\pi$.
An effective criterion for the stability of solutions to Hill's equation is $\dl(\la)\in\mathbb{R}$ and either $|\dl(\la)|<2$
or $|\dl(\la)|=2$ and $M(\la)\in\mathbb{Z}_{2}$. From eq. \eqref{monodromy} it follows that $\dl(\la) = y_{1}(\pi;\la)+y'_{2}(\pi;\la)$.
Therefore, solving $\dl(\la) =\pm 2$ for $\la$ we obtain segments on $\mathbb{R}_{+}$ where $|\dl(\la)|<2$ as well as those,
where $|\dl(\la)|>2$. The former are termed bands whereas the latter -- gaps. Thus, the spectrum 
of Hill's equation \eqref{HillEq} has the band structure. This fact is a thesis of 
\begin{quotation}
\noindent
\textbf{Theorem}(Oscillation) Let $\la\in\mathbb{C}$ and $\underline{\la}_{i},\overline{\la}_{i}\in \mathbb{R}_{+}$ 
such that $\dl(\underline{\la}_{i})=2$ for $i\in\mathbb{N}_{0}$ 
and $\dl(\overline{\la}_{i}) = - 2$ for $i\in\mathbb{N}$. Then solutions to eq. \eqref{HillEq} are stable
if 
$$
\la\in \bigcup_{i\in\mathbb{N}} (\underline{\la}_{2i-2},\overline{\la}_{2i-1}) \cup (\overline{\la}_{2i},\underline{\la}_{2i-1}),
$$
and in case when $\la = \overline{\la}_{2i-1} = \overline{\la}_{2i}$ 
or $\la = \underline{\la}_{2i-1} = \underline{\la}_{2i}$ for $i\in I \subset \mathbb{N}$ and unstable otherwise.
\end{quotation}

%
%

\providecommand{\href}[2]{#2}\begingroup\raggedright\endgroup

\end{document}